\documentclass[%
 aip,
 amsmath,amssymb,
 reprint,%
]{revtex4-1}

\usepackage{graphicx}
\usepackage{dcolumn}
\usepackage{bm}

\usepackage[utf8]{inputenc}
\usepackage[T1]{fontenc}
\usepackage{mathptmx}
\usepackage{etoolbox}
\usepackage{amssymb, epsfig,amssymb, latexsym}
\usepackage{amsfonts,psfrag,amsmath,bbm,color,url}
\usepackage{marvosym}
\usepackage{titlesec}
\usepackage{comment}
\usepackage{soul}

\usepackage{caption,subcaption}
\usepackage{comment}

\usepackage{multirow}

\def\PP{{{\rm l}\kern - .15em {\rm P} }}
\def\PN2{{\PP_{N}-\PP_{N-2}}}

\newcommand{\deleted}[1]{{}}

\usepackage{tikz}
\usetikzlibrary{fit,positioning}
\usepackage{siunitx}
\usepackage{float}
\makeatletter
\def\@email#1#2{%
 \endgroup
 \patchcmd{\titleblock@produce}
  {\frontmatter@RRAPformat}
  {\frontmatter@RRAPformat{\produce@RRAP{*#1\href{mailto:#2}{#2}}}\frontmatter@RRAPformat}
  {}{}
}%
\makeatother
\begin{document}

\preprint{AIP/123-QED}

\title[A Stochastic Precipitating Quasi-Geostrophic Model]{A Stochastic Precipitating Quasi-Geostrophic Model}
\author{Nan Chen}
\author{Changhong Mou}%
\email{mouc@purdue.edu}
\author{Leslie M. Smith}

\affiliation{ 
Department of Mathematics, University of Wisconsin-Madison, Madison, WI, USA
}%
\author{Yeyu Zhang}
\affiliation{%
School of Mathematics, Shanghai University of Finance and Economics, Shanghai, China
}%
\date{\today}

\begin{abstract}
Efficient and effective modeling of complex systems, incorporating cloud physics and precipitation, is essential for accurate climate modeling and forecasting. However, simulating these systems is computationally demanding since microphysics has crucial contributions to the dynamics of moisture and precipitation. In this paper, appropriate stochastic models are developed for the phase-transition dynamics of water, focusing on the precipitating quasi-geostrophic (PQG) model as a prototype. By treating the moisture, phase transitions, and latent heat release as integral components of the system, the PQG model constitutes a set of partial differential equations (PDEs) that involve Heaviside nonlinearities due to phase changes of water. Despite systematically characterizing the precipitation physics, expensive iterative algorithms are needed to find a PDE inversion at each numerical integration time step. As a crucial step toward building an effective stochastic model, a computationally efficient Markov jump process is designed to randomly simulate transitions between saturated and unsaturated states that avoids using the expensive iterative solver. The transition rates, which are deterministic, are derived from the physical fields, guaranteeing physical and statistical consistency with nature. Furthermore, to maintain the consistent spatial pattern of precipitation, the stochastic model incorporates an adaptive parameterization that automatically adjusts the transitions based on spatial information. Numerical tests show the stochastic model retains critical properties of the original PQG system while significantly reducing computational demands. It accurately captures observed precipitation patterns, including the spatial distribution and temporal variability of rainfall, alongside reproducing essential dynamic features such as potential vorticity fields and zonal mean flows.
\end{abstract}

\maketitle

\section{Introduction}

Over the past decade, numerous extreme weather events have occurred continuously in subtropical and temperate regions, \cite{winters2014role, houze2017multiscale, sapsis2021statistics, vitart2018sub, santoso2017defining, chekroun2024high, lucarini2016extremes}. These extreme weather events are often closely related to the Earth's hydrological cycle \cite{newman2012relative, lavers2015climate}. Therefore, a better understanding of moisture, cloud physics, and an accurate representation of water in different phases will undoubtedly enhance our ability to mathematically model these extreme events, deepen our physical understanding, and help meteorologists make more accurate predictions and judgments about them.

In the atmosphere, the physical process of phase transitions of water involves a wide range of spatial and temporal scales, from cloud condensation nuclei (in millimeters and on the order of seconds) to the formation of cloud clusters and superclusters spanning thousands of kilometers and lasting hours to days. Thus, directly utilizing full-order models (FOM) for modeling, understanding, and predicting such complex multiscale physical systems is by no means an easy task, and their large scale and complex structure will also lead to almost insurmountable mathematical and computational challenges \cite{majda2012challenges, majda2018model}.

Reduced-order models (ROMs) provide valuable tools for understanding and predicting the weather and climate systems \cite{palmer2001nonlinear, majda2008applied, edwards2011history, franzke2005low, mou2023efficient}. In particular, stochastic models have become appealing alternatives for extended-range prediction and climate sensitivity studies due to the chaotic nature of these systems \cite{majda2001mathematical, franzke2015stochastic, imkeller2001stochastic, majda1999models, majda2003systematic, chen2018conditional}. Stochastic models are more computationally efficient than the state-of-the-art climate models. They effectively reproduce low-frequency dynamical processes and have demonstrated comparable predictive skills at large scales \cite{palmer2019stochastic, horenko2008automated, majda2009normal}. In the standard procedure of developing a stochastic model for large-scale dynamics, the small-scale features are often replaced by simple random noise. However, the dynamics of moisture and precipitation strongly rely on the details of the microphysics, which brings new challenges when developing effective stochastic models for such complex systems.

The objective of this study is to develop appropriate stochastic models for phase-transition dynamics of water, focusing on the precipitating quasi-geostrophic (PQG) model~\cite{smith2017precipitating} as a prototype. For preliminary inquiries, we confine our attention to the two-level PQG equations, deferring the exploration of three-dimensional PQG dynamics to subsequent research. The new stochastic precipitating quasi-geostrophic (SPQG) model aims to advance the understanding and forecasting of atmospheric dynamics in moist mid-latitude regions~\cite{law2015data,asch2016data}.

The PQG model builds upon the classical quasi-geostrophic (QG) model~\cite{charney1948scale}, the latter which is an important achievement in atmospheric dynamics for characterizing incompressible fluids under the conditions of rapid rotation and strong stratification, pertinent to description of the troposphere at synoptic scales~\cite{majda2003introduction, vallis2017atmospheric}. The traditional QG model does not account for moisture, which is critical in atmospheric dynamics, as water facilitates key energy transfer mechanisms in the atmosphere. This omission necessitates the exploration of moist variants of the QG system. It is worth highlighting that a distinctive feature of the PQG model~\cite{smith2017precipitating} is its incorporation of moisture, phase transitions, and latent heat release as integral components of the system. This contrasts with other moist models that typically treat these factors as supplementary elements in a more ad hoc fashion~\cite{emanuel1987baroclinic, lapeyre2004role}. Consequently, the PQG equations constitute a set of partial differential equations (PDEs) that involve Heaviside nonlinearities due to phase changes of water (between phases of vapor, liquid, etc.).

In the dry QG model, the dynamics are reduced to a single scalar variable, the potential vorticity (PV)~\cite{ertel1942neuer}. The entire QG dynamics are expressed in terms of PV, which requires the inversion of a rescaled Laplacian, a process known as potential vorticity inversion or PV inversion~\cite{hoskins1985use}. When moisture is included in the PQG model, the dynamics are described by two variables: an equivalent PV, which slightly differs from the dry PV, and a moist scalar variable $M$. This necessitates PV-and-M inversion, which involves the inversion of a more complex operator. This complexity mainly involves a nonlinear elliptic PDE with discontinuous coefficients due to phase transitions. During numerical simulations, dealing with this nonlinear elliptic PDE is challenging because the coefficients depend on phase boundaries, which are not known beforehand. Therefore, iterative algorithms are required to find the convergent solution to the elliptic equation at each numerical integration time step~\cite{smith2017precipitating, edwards2020spectra, hu2021initial, mou2023combining}. Once convergence is achieved, the phase boundaries can be identified as part of the inversion process. However, the high degrees of freedom in spatial discretization and the nonlinear nature of this elliptic equation typically result in significant computational demands.

To address the challenge above, the construction of the new stochastic precipitating quasi-geostrophic (SPQG) model focuses on managing the complexity arising from a nonlinear elliptic PDE due to phase changes. This approach aims to provide a more efficient modeling and computational tool. With this motivation, the stochastic surrogate model developed in this paper replaces the nonlinear elliptic equation with Markov jump processes \cite{gardiner2004handbook, katsoulakis2003coarse, khouider2010stochastic} to simulate the phase transitions of water molecules from vapor to liquid. Stochastic processes are employed to optimize the identification of phase boundaries. Specifically, the Markov jump process is used to randomly simulate transitions between saturated and unsaturated states at a given location. A saturated state indicates precipitation, while an unsaturated state indicates no precipitation. It is worth noting that the likelihoods of these random transitions are determined by transition rates derived from PV fields. They are deterministic and physics-driven. In addition to significantly lowering computational costs, the development of the SPQG model also focuses on optimizing physical realism by maintaining essential cloud physics and the natural physical characteristics inherent in the PQG system. An adaptive stochastic parameterization method is designed, automatically adjusting the transitions based on spatial information to ensure the cloud fraction estimation remains within reasonable physical bounds. This approach helps maintain consistent spatial patterns of precipitation.

The structure of the remainder of the paper is outlined as follows. Section~\ref{ss-pqg} introduces the PQG equations. In Section~\ref{ss-spqg}, we present the stochastic surrogate models for the PQG equations. Numerical results are presented in Section~\ref{ss-spqg-numerical}, followed by additional discussions and concluding remarks in Section~\ref{ss-conclusion}.

\section{The Precipitating Quasi-Geostropic Equations\label{ss-pqg}}

This section starts by reviewing the classical quasi-geostrophic (QG) model (namely, the dry QG model), which is primarily characterized by a single forecast variable: the potential vortcity (PV). The details of this model have been thoroughly documented in seminal works such as \cite{salmon1980baroclinic, qi2016low, vallis2017atmospheric}. In the dry QG model, the inversion of PV is given by a simple linear elliptic equation. Building upon the dry QG model, the PQG equations will be introduced. The PQG model includes the additional moisture variable $M$ \cite{smith2017precipitating}, and requires a more comprehensive formulation for the nonlinear PV-and-M inversion. The general procedure of developing stochastic models will be based on the two-level PQG equations. This model preserves the essential dynamical characteristics of the three-dimensional PQG framework, incorporating both the height-independent barotropic modes and the baroclinic modes. A brief overview of the numerical simulation results for this model will be presented, demonstrating new behaviors in PQG turbulence due to moisture and phase changes, as well as potential physical characteristics and correlations in the turbulent field.

\subsection{The classical QG equations}

The QG model is derived as a limit of the Boussinesq dynamical model for rotating and stratified fluids.
Here all variables, equations and parameters are presented in non-dimensional form.
The model is constructed to represent slowly evolving winds and temperature at length scales $L \sim 1000$ km and times scales $T \sim$ days, for which the earth's rotation and stratification dominate the dynamics. Mathematically, the derivation assumes rapid rotation (small $Ro$) and strong stable stratification (small $Fr$) in the distinguished limit $Fr \sim Ro = \epsilon \rightarrow 0$. The non-dimensional numbers are defined as $Ro = U/(fL)$ and $Fr = U/(NH)$, where $U$ is the reference horizontal velocity, $H$ is the reference height, $f$ is the Coriolis parameter, and $N$ is the buoyancy frequency. Within this framework, QG centers on the scalar quantity PV, defined as \( PV = \xi + F\partial_z \theta \), where \( \xi = \partial_x v - \partial_y u \) is the vertical component of relative vorticity, $\theta$ is the potential temperature, and the parameter $F=Fr/Ro = O(1)$.  As constructed, PV integrates information about the horizontal winds \( \mathbf{u}_h = (u, v) \), potential temperature \( \theta \), and pressure \( \psi \), and it is a slowly varying quantity that evolves independently from inertia-gravity waves.

The QG system follows from three key features of the limiting dynamics. First, rapid rotation leads to geostrophic balance between the Coriolis acceleration and horizontal pressure gradients, such that \( \mathbf{u}_h = (-\partial_y \psi, \partial_x \psi) \), where $\psi$ is the pressure. Second, strong stable stratification implies the hydrostatic balance \( \theta = F \partial_z \psi \), which is the balance between the buoyancy $\theta$ and the vertical temperature gradient (appropriately scaled).
Third, a solvability constraint on the dynamics requires that PV is a materially conserved quantity, and its advection is described by
\begin{equation}
\partial_t PV + \mathbf{u}_h \cdot \nabla_h PV = 0,
\label{eqn:PVevolution}
\end{equation}
where $\nabla_h = (\hat{\bf x}\; \partial_x, \hat{\bf y} \;\partial_y)$.

Using the definition of PV together with the geostrophic and hydrostatic balance relations, one arrives at the elliptic partial differential equation
\begin{equation}
     PV = \Delta_h \psi + F^2 \partial^2_{z} \psi. \label{eqn:PV-inversion}
\end{equation}
This linear elliptic equation is the basis for numerical integration of the PV-evolution equation \eqref{eqn:PVevolution}, as follows. Given an initial ${\bf u}_h$ and PV, one may evolve PV in time using \eqref{eqn:PVevolution}.  Subsequently, \eqref{eqn:PV-inversion} may be inverted to find the pressure $\psi$, and then the balance relations determine ${\bf u}_h$ and $\theta$, such that updated PV and ${\bf u}_h$ are available for the next evolution step.

\subsection{The three-dimensional PQG equations}
\label{sec:3D-PQG}
The precipitating quasi-geostrophic (PQG) model~\cite{smith2017precipitating, edwards2020spectra, hu2021initial} transcends the conventional quasi-geostrophic dynamics by incorporating the additional complexity of moisture, including the dynamics of water, phase transitions, and precipitation. Moisture is represented by a scalar variable \( q_t \)~\cite{hernandez2013minimal}, denoting the total water mixing ratio, with saturation assumed at \( q_t = q_{vs} \). One may introduce the definitions \( H_u =  \mathbbm{1}(q_t < q_{vs}) \) and \( H_s = 1 - H_u = \mathbbm{1}(q_t \geq q_{vs}) \) to distinguish between unsaturated (\( q_t < q_{vs} \)) and saturated (\( q_t \geq q_{vs} \)) environments.
In unsaturated regions, water is in vapor phase $q_t=q_v$, while in saturated environments, water contains vapor at saturation $q_v = q_{vs}$ together with rain water $q_r$, such that $q_t = q_{vs} + q_r$.
Thus, one may find $q_v, q_r$ from $q_t, q_{vs}$ using the relations $q_v =  {\rm min}(q_t,q_{vs})$ and $q_r = {\rm max}(0,q_t - q_{vs}).$

The derivation of the PQG equations follows the same ideas as that of the dry QG equations.  The main assumptions are rapid rotation and strong stable stratification, where the latter assumption is applied to {\it both} temperature {\it and} water. In mathematical terms, these assumptions imply the distinguished limit $Fr_s \sim Fr_u \sim Ro = \epsilon \rightarrow 0$, where $Fr_s$ (for saturated regions) and $Fr_u$ (for unsaturated regions) are Froude numbers defined as $Fr_s = U/(N_s H)$ and $Fr_u = U/(N_u H)$. Here, $N_s$ and $N_u$ denote the buoyancy frequencies for saturated and unsaturated regions, respectively, analogous to the buoyancy frequency $N$ of a dry system. A physical constraint is $Fr_s > Fr_u$.

With the inclusion of moisture into Boussinesq dynamics, geostrophic balance remains unchanged. However, moist-Boussinesq hydrostatic balance reflects information about phase changes between unsaturated and saturated fluid parcels, because the buoyancy includes contributions from dry air $\theta$, water vapor $q_v$ and liquid water $q_r$, and the total buoyancy changes its functional form in unsaturated and saturated regions. Since phase changes are fast processes, a good definition of PV is delicate, since ideally it should be a slowly varying quantity representative of the the large scales $L \sim 1000$ km and slow times $T \sim$ days.  For instance, although potential temperature $\theta$ dominates the buoyancy in the limiting dynamics, one can see that $\theta$ is directly impacted by fast phase transitions because its evolution equation is forced by terms representing latent heat exchange between vapor and liquid (condensation and evaporation terms).  Thus, using $\theta$ to define PV is problematic in the sense that such a PV is not slowly varying in the moist dynamics.  However, a definition of {\it moist} PV based on the {\it equivalent} potential temperature $\theta_e = \theta + q_v$ provides a slowly varying moist ${\rm PV}_e$. The quantities $\theta$ and $q_v$ are forced by oppositely-signed latent-heat terms, and thus evolution of their sum $\theta_e$ is free of fast, oscillatory forcing terms. Consequently, the definition of moist $PV_e = \xi + {\cal F} \theta_e$ is a slowly varying quantity, in direct analogy with the dry PV.  An additional variable $M = q_t + {\cal G}\theta_e$ is necessary to fully describe the slowly varying, moist dynamics \cite{smith2017precipitating}.  The constants
${\cal F} = Fr_s/Ro$, and ${\cal G} =
Fr_s (Fr^{-2}_u - Fr^{-2}_s)^{1/2}$ are $O(1)$ for consistency with the PQG limiting parameter regime.

The governing equations of the PQG model are given by \cite{smith2017precipitating},
\begin{equation}
\frac{D_h PV_e}{Dt} =  -{\cal{F}}\frac{\partial \mathbf{u}_h}{\partial z} \cdot\nabla_h\theta_e,
\label{eqn:PQG-PV-budget}
\end{equation}
\begin{equation}
\frac{D_h M}{Dt} = V_r\frac{\partial q_r}{\partial z},
\label{eqn:PQG-M}
\end{equation}
where $\frac{D_h}{Dt}$ is the horizontal material derivative, defined as $\frac{D_h}{Dt} (\cdot) = \partial_t(\cdot) + \mathbf{u}_h \cdot \nabla_h$ and $V_r$ is the non-dimensional rainfall speed. In order to use \eqref{eqn:PQG-PV-budget}--\eqref{eqn:PQG-M}, we must close the system using the definitions of ${\rm PV}_e$ and $M$ together wtih geostrophic and hydrostatic balance.  These relations
lead to a nonlinear elliptic PDE 
\cite{zhang2021effects}:
\begin{gather}
\nabla_h^2 \psi + \frac{\partial}{\partial z} \left[ \dfrac{1}{1+\cal{G}}H_u\left( {\cal{F}}^2\frac{\partial \psi}{\partial z} +  {\cal{F}}M\right)\right] +\\ \frac{\partial}{\partial z} \left[H_s\left( {\cal{F}}^2\frac{\partial \psi}{\partial z} + {\cal{F}}
q_{vs}\right)\right] = PV_e,
\label{eqn:PV-M-inversion}
\end{gather}
where $\psi$ is the associated pressure.
The existence and uniqueness of solutions to
\eqref{eqn:PV-M-inversion} is proved in
\cite{remond2024nonlinear}.
The winds, temperature and water fields are then determined by
\begin{gather}
\mathbf{u}_h = \left (-\frac{\partial \psi}{\partial y}, \frac{\partial \psi}{\partial x} \right ),\quad q_t = M - {\cal{G}}\theta_e,\\
\theta_e =\dfrac{1}{1+\cal{G}} H_u\left ( {\cal{F}}\frac{\partial \psi}{\partial z}+ M \right ) + H_s\left ({\cal{F}} \frac{\partial \psi}{\partial z} + q_{vs} \right ).
\label{eqn:slowuthetae}
\nonumber
\end{gather}

\medskip
\noindent\textbf{Major computational challenge.} The PDE \eqref{eqn:PV-M-inversion} is nonlinear, as discussed in \cite{smith2017precipitating, hu2021initial}, due to the dependence of the Heaviside functions $H_u$ and $H_s$ on the quantity $q_t$, which in turn depends on $\psi$ as seen in (\ref{eqn:slowuthetae}). Therefore, an expensive iterative scheme is required for the solution of this nonlinear PDE. 

In summary, unlike the dry model, which requires solving a linear elliptic PDE (\ref{eqn:PV-inversion}) during the inversion phase, the moist model necessitates the inversion of a nonlinear elliptic PDE (\ref{eqn:PV-M-inversion}) with discontinuous coefficients given by the Heaviside functions. This introduces additional complexity associated with the iterative process, culminating in elevated computational demands. Therefore, the main objective of this paper, introduced in Section~\ref{ss-spqg}, is to present a stochastic precipitating quasigeostrophic (SPQG) model. The SPQG model aims to replace the solver of the computationally intensive elliptic equations in the traditional PQG model with more efficient stochastic processes. This approach seeks to reduce computational costs while retaining the essential precipitation and cloud physics from the PQG model.

\subsection{Two-level PQG equations}
\label{ss:two-level-PQG}

The physical phenomena of the PQG system have been analyzed through both analytical and numerical studies, concentrating on specific solution classes relevant to meteorological applications \cite{wetzel2017moisture, wetzel2019discontinuous, edwards2020atmospheric}. Broader studies of the PQG equations have been limited to numerical methods \cite{edwards2020spectra, hu2021initial}. Notably, both numerical studies \cite{edwards2020spectra, hu2021initial} investigate the two-level PQG system, which simplifies vertical dependence to two horizontal layers. Nevertheless, the inversion ${\rm PV}_e$ (henceforth abbreviated PV) and $M$ remains the most computationally demanding aspect of solving the two-level PQG system, as demonstrated in~\cite{hu2021initial}. In developing a general procedure for the SPQG model, aiming to replace the computationally intensive elliptic equations with more efficient stochastic processes, we will focus on the two-level PQG equations.

Analogous to the approach employed for the dry two-level QG system, the PQG system is deduced by applying a staggered vertical grid \cite{qi2016low,hu2021initial}. In the two-level model, the subscript $(\cdot)_{j}$, where $j=1,2$, signifies variables assigned to either the first or second level, while the subscript $(\cdot)_{m}$ denotes variables situated at the intermediate level due to the use of the staggered vertical grid. Note that the thermodynamic variables, namely $M$, $q_t$, and $q_r$, are defined in the intermediate layer where they interact with PV and $\psi$ within a simplified two-level system, avoiding boundary dynamics. Therefore, it is more appropriate to consider the intermediate quantities of PV and $\psi$, which are interpolated between layer 1 and layer 2. The two-level equations are obtained from~\eqref{eqn:PQG-PV-budget}--\eqref{eqn:PQG-M} using centered finite differences in \(z\) and a rigid lid boundary condition \(w = 0\) as described in~\cite{hu2021initial}. The two-level version of the PQG equations is then given by:
\begin{equation}
\frac{D_1 PV_1}{Dt} =  - {\cal{F}}\frac{\partial \mathbf{u}_h}{\partial z} \cdot\nabla_h\theta_{e,1},
\label{eqn:PQG-layer-1}
\end{equation}
\begin{equation}
\frac{D_2 PV_2}{Dt} =  -{\cal{F}}\frac{\partial \mathbf{u}_h}{\partial z} \cdot\nabla_h\theta_{e,2},
\label{eqn:PQG-layer-2}
\end{equation}
\begin{equation}
\frac{D_m M_m}{Dt} =V_r \frac{\partial q_{r}}{\partial z},
\label{eqn:PQG-layer-3}
\end{equation}
where the subscripts 1, 2, and $m$ denote the material derivative at each level:
$\frac{D_1}{Dt}(\cdot) = \partial_t(\cdot) + u_1 \partial_x(\cdot) + v_1 \partial_y(\cdot)$, and similarly for \(\frac{D_2}{Dt}(\cdot)\) and \(\frac{D_m}{Dt}(\cdot)\). Note that the terms \(\partial \mathbf{u}_h / \partial z\) and \(\partial q_r / \partial z\) are expressed as functions of a continuous \(z\) coordinate to simplify notation. For more details, please refer to~\cite{hu2021initial}.

To complete the two-level PQG system, it is necessary to perform the PV-and-M inversion to determine the remaining meteorological fields. Additionally, for numerical simulations, some features are added to the dynamical model in~\eqref{eqn:PQG-layer-1}--\eqref{eqn:PQG-layer-3}, such as an evaporation source term, lower-level friction, and background velocity and temperature to provide conditions for baroclinic instability. To maintain the clarity and concise structure of the article, we will not elaborate on these details here. Instead, we have included them in Appendix~\ref{ap:A} for reference, starting with the equations, which include detailed parameter definitions and additional physical settings.

To complete the two-level PQG system, it is necessary to perform the PV-and-M inversion to determine the remaining meteorological fields. Additionally, for numerical simulations, some features are added to the dynamical model in~\eqref{eqn:PQG-layer-1}--\eqref{eqn:PQG-layer-3}, such as an evaporation source term, lower-level friction, and nonzero background velocity and temperature to provide conditions for baroclinic instability.  To maintain the clarity and concise structure of the article, we will not elaborate on these details here. Instead, we have included them in Appendix~\ref{ap:A} for reference, starting with the equations, which include detailed parameter definitions and additional physical settings.

\subsection{Physical features in the PQG turbulence} \label{into:PV-and-Cloud}

This subsection aims to provide brief numerical simulation results of the PQG dynamics. The goal is to highlight the new features that arise in the PQG turbulence due to moisture and phase changes, as well as some 
physical characteristics of turbulent fields and their correlations.

Building on the two-level PQG model discussed in Section~\ref{ss:two-level-PQG} as the simplest setup to include the effects of moisture and phase changes, the simulations are conducted using standard pseudospectral methods to stay in close alignment with traditional numerical methods for turbulent fluid dynamics. The parameters selected for our numerical experiments include the evaporation rate \( E \) and the rainfall speed \( V_r \), with detailed descriptions in Appendix~\ref{ap:A}, and other parameters settings provided in Section~\ref{ss-spqg-parameter}. Our discussion will primarily focus on the standard values of \( E = 0.2 \), \( V_r = 1.0 \), and the saturation mixing ratio $q_{vs}$, which will be dynamically determined by equation (\ref{def:qvs-2level}) in Appendix~\ref{ap:A} with the parameter settings \( q^0_{vs} = q^1_{vs} = 1.0 \). These values will remain the default unless otherwise specified.

\begin{figure*}
    \centering
\includegraphics[width=.8\textwidth]{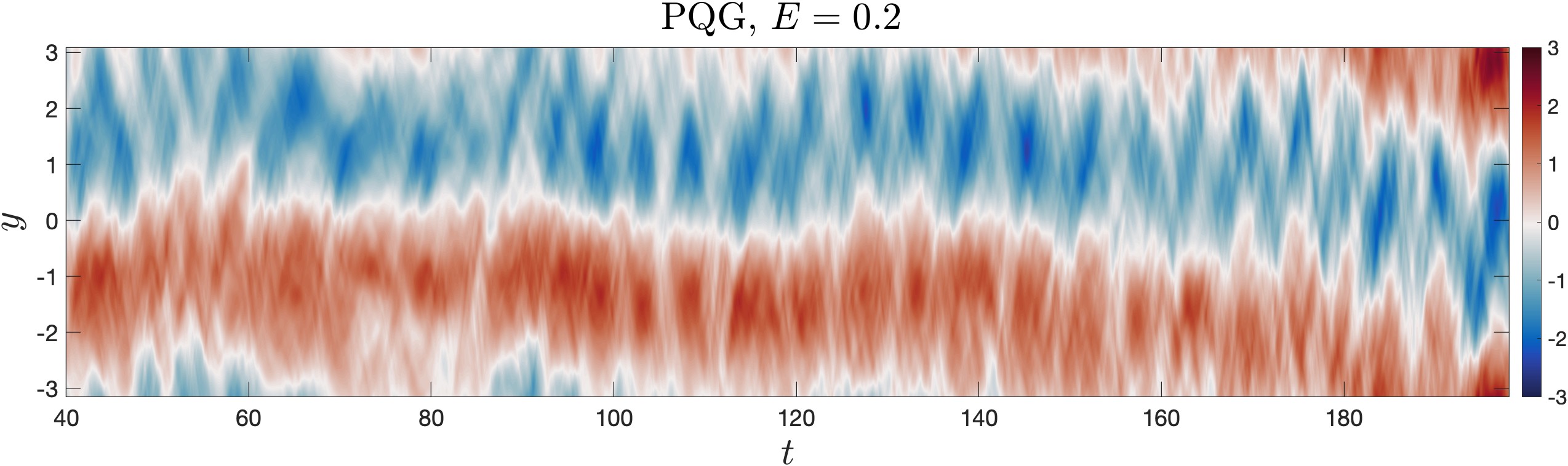}
\includegraphics[width=.8\textwidth]{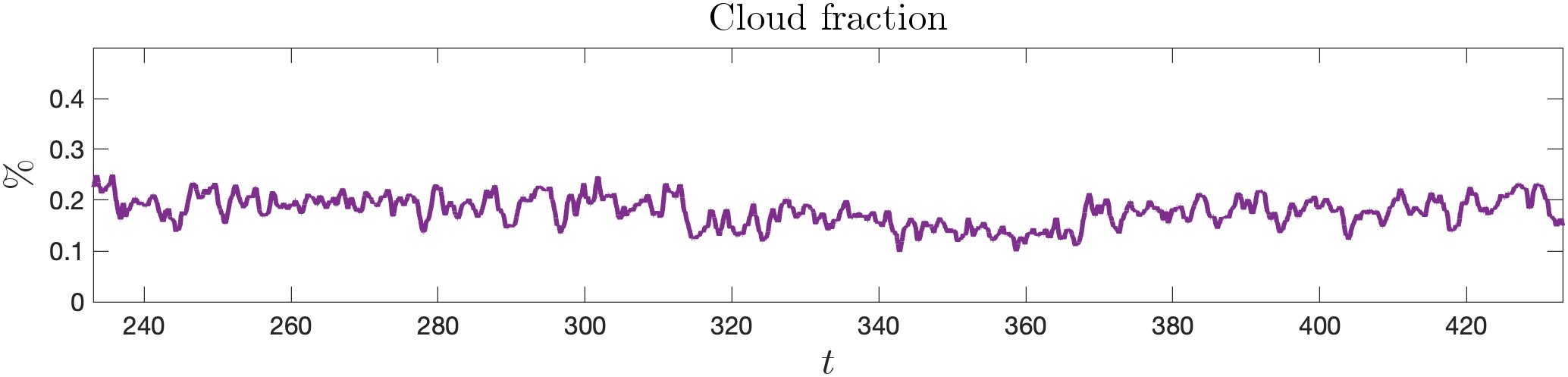}
\includegraphics[width=.8\textwidth]{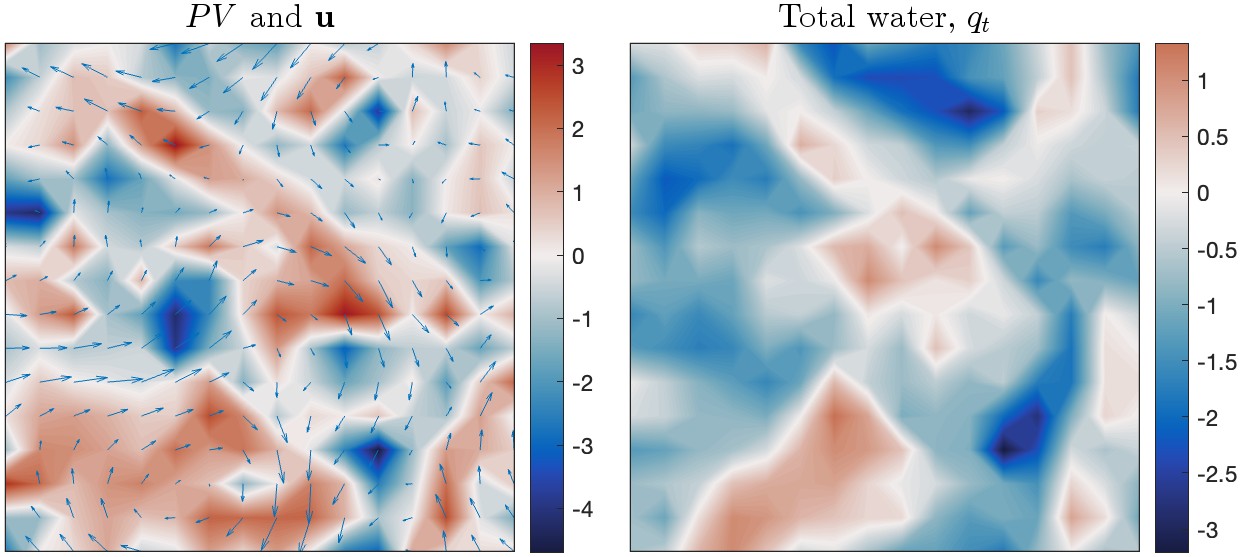}
    \caption{The physical features in the PQG turbulence include the zonal jet (top panel), cloud fraction (middle panel), and snapshots of the PV along with the velocity field \(\mathbf{u}_h\) represented by blue arrows (bottom left), and the total water content \(q_t\) (bottom right).  All spatially 2D plots are fixed-time, $(x,y)$-slices of fields at the intermediate height between levels 1 and 2, where $x$ $(y)$ is the zonal (meridional) direction.  variables are defined at the intermediate level.  Also note that, unless otherwise specified, all plots related to the \(q_t\) quantity in this paper have been adjusted by subtracting the saturation mixing ratio \(q_{vs}\). Consequently, in the \(q_t\) plots, values above 0 (red areas) will indicate rainwater.}
     \label{fig:pqg-evap-3-a}
\end{figure*}

A signature of QG flows is the development of zonal jets, with jet features that vary depending on parameters such as the background wind and the $\beta$-parameter characterizing the change in rotation rate with latitude. \cite{qi2016low,vallis2017atmospheric,treguier1987oceanic}. 
For our parameter regime relevant for midlatitude flows with precipitation,
Figure~\ref{fig:pqg-evap-3-a} (top panel) displays the upper-level, zonally-averaged wind as a function of time, from $t=40$ to $t=200$. We remind the reader that these winds ${\bf u}$ are $2\pi$-periodic fluctuations from a background shear profile $(-U,U)$ in the zonal direction. One can see distinct eastward (red) and westward (blue) jets, with both high- and low-frequency variability in time.  The reconstruction of gross statistical features of the jets is crucial for the development of a surrogate model, for instance, key characteristics are the width of the jets, as well their temporal mean and variance.
The middle panel of Figure~\ref{fig:pqg-evap-3-a} demonstrates the temporal variation in cloud fraction associated with the jet stream observed above. The cloud fraction is calculated as the $L_1$ norm of the cloud indicator \( H_s =  \mathbbm{1}(q_t \geq  q_{vs}) \).

The bottom panel of Figure~\ref{fig:pqg-evap-3-a} provides information on the PV, the velocity field \(\mathbf{u}\), and the total water content \(q_t\), highlighting a potential correlation between these variables. Here, the quantity PV is defined as the value at the intermediate level, namely $PV:=(PV_1+PV_2)/2$. This relationship is such that regions with high PV values exhibit a tendency toward cyclonic motions. At the same time, these regions correspond to relatively high water content, indicating the presence of more rainwater and precipitation. Similarly, regions with low PV exhibit a tendency toward anticyclonic motions and low water content. Note that the PV often serves as a powerful dynamical tracer and a formidable instrument for describing mid-latitude dynamics~\cite{hoskins1985use}. Recent research findings also suggest that PV is closely related to phase transitions, precipitation, and deep moist convection in extreme events on medium and small scales~\cite{martius2006episodes, roberts2000relationship, chagnon2009horizontal}. Therefore, in Sections~\ref{ss-spqg}, within the SPQG model design of the Markov jump process, we aim to integrate data and utilize this potential correlation to enhance the mathematical precision and retain the inherent physics of the model.

\begin{table}[H]
    \centering
    \begin{tabular}{c|c}
         Evaporation rate $E$ & Cloud fraction  $\bar{C}$
         \\
         \hline\hline
         0.02& $3.76\%$ \\
         0.1& $8.40\%$ \\ 
         0.2& $ 25.87\%$ \\
         0.3& $ 32.02\%$\\  
         0.35& $ 38.64\%$ \\
         0.5& $92.55\%$ 
    \end{tabular}
    \caption{Different mean cloud fractions for different evaporation rates in PQG
    }
    \label{tab-cloud}
\end{table}

It is worth noting that in the PQG system, the moisture impact can be modulated by tuning the evaporation rate \( E \). The mean cloud fraction, which is calculated as follows, is listed in Table~\ref{tab-cloud}:
\begin{align}
    \bar{C} = \frac{1}{N}\sum_{k=1}^N C(t_k),
\end{align}
where \( N \) is the number of snapshots, and \( C(t_k) \) is the cloud fraction at time \( t_k \). A larger evaporation rate usually results in a higher mean cloud fraction, thus a richer rainwater content. In alternative parameter regimes characterized by a reduced or increased cloud fraction, patterns analogous to those depicted in Figure~\ref{fig:pqg-evap-3-a} are preserved, including the meridional propagation of zonal jets and a correlation between the PV fields and the total water content \( q_t \). Therefore, in the subsequent SPQG numerical simulation experiments, we will primarily discuss the typical scenario where \( E = 0.2 \) and compare it with the numerical results of the high-resolution PQG model discussed in this section.

\section{Developing a Stochastic Precipitating Quasi-Geostropic (SPQG) Model \label{ss-spqg}}

\subsection{Overview of the stochastic model \label{ss-stochastic}}
In the PQG framework, solving the stream functions, which are the essential step toward determining the locations of precipitation, requires an extensive iterative PV-and-M inversion process. More specifically, running the governing equations of the PVs and the $M$ variables in \eqref{eqn:layer-pqg-1}--\eqref{eqn:layer-pqg-3} requires the input from the streamfunctions $\psi_1$ and $\psi_2$. However, these streamfunctions need to be solved from the nonlinear elliptic equations \eqref{eqn:pv-m-inversion-1}--\eqref{eqn:pv-m-inversion-2}, where the Heaviside functions $H_s$ and $H_u$ depends nonlinearity on the total water mixing ratio $q_t$ that is a highly nonlinear function of the streamfunctions. Therefore, an expensive iterative scheme is required to resolve this nonlinear PDE. As the spatiotemporal patterns of the solution exhibit strong turbulent behavior, a natural way to reduce the computational cost is to develop a stochastic model. It aims to reproduce the turbulent features using random processes and find an efficient approximation solution of the PV-and-M inversion without using the expensive iterative method. As in the PQG model, the variable PV is defined as the PV value at the intermediate level, namely $PV:=(PV_1+PV_2)/2$. See Appendix~\ref{ap:A}.

\subsubsection{Improving the computational efficiency of solving the elliptic equation using stochastic processes}
\noindent\emph{(a) Modeling Heaviside nonlinearity using Markov jump process.}

The SPQG equations introduce a new approach by replacing the iterative method for solving the nonlinear elliptic equations of streamfunctions, detailed from equations \eqref{eqn:pv-m-inversion-1} to \eqref{eqn:pv-m-inversion-2}, with a much cheaper solver involving the use of a Markov jump process \cite{gardiner2004handbook} at each mesh grid. Since precipitation often occurs randomly at discrete time intervals, the Markov jump process is a natural choice to effectively characterize the state changes. Denote by $\mathcal{X}_t(x,y)$ the Markov jump process at location $(x,y)\in \Omega$ and time $t$, where $\Omega$ is the solution domain. The variable $\mathcal{X}_t(x,y)$ undergoes random transitions between two distinct states: the saturated state $s_t$ and the unsaturated state $u_t$. It is used to replace $H_s$ in \eqref{eqn:pv-m-inversion-1} to \eqref{eqn:pv-m-inversion-2}, where $H_u$ becomes $1-\mathcal{X}_t(x,y)$ at each grid point. When these Heaviside functions are described by Markov jump processes, the evaluation of $q_r$ from the stream function and other variables is avoided. Therefore, expensive iterative methods are no longer needed and the solution of the stream functions in \eqref{eqn:pv-m-inversion-1} to \eqref{eqn:pv-m-inversion-2} can be easily obtained by solving the resulting linear elliptic equations. Although the Markov jump process is stochastic, the transition rates that describe the probability of changing the states are deterministic. The transition rates depend on the instantaneous physical fields such that the spatiotemporal solution of the Markov jump processes to be correlated. This allows the resulting solution to capture large-scale coherent structures and small-scale turbulent features. One crucial feature of the proposed Markov jump processes is that, despite the intrinsic spatial correlation, each Markov jump process in $\Omega$ can operate independently, reducing the computational cost compared to solving the original elliptic equations.\\

\noindent\emph{(b) Determining the transition rates in the Markov jump process.}

Within the framework of the Markov jump process, the transition rate plays a pivotal role. It defines the probability of the system transitioning from one state to another within a specified time frame, effectively quantifying the frequency of these transitions. It is observed that there is a consistent movement between the PV and the cloud coverage (see Section~\ref{ss-pqg}). In fact, in dry dynamics, PV is already conserved under frictionless and adiabatic conditions~\cite{schubert2004english}. This conservation property renders PV a potent dynamical tracer and provides a robust framework for characterizing the dynamics of mid-latitudes~\cite{hoskins1985use}. Later, the PV is also found to be intricately linked with phase transitions, precipitation, and deep moist convection during extreme weather events at medium and small scales, as observed in studies by ~\cite{martius2006episodes,roberts2000relationship,chagnon2009horizontal}. These correlations offer a valuable intuition for employing the PV information as a predictive indicator for determining transition rates and for informing the distribution of water. Therefore, the SPQG model incorporates a mechanism where regions with weaker PV values correspond to relatively lower transition rates, while stronger PV values indicate higher transition rates. On the other hand, the cloud coverage is not uniquely determined by the PV. The areas characterized by positive PV constitute about 50\% of the observed field, and cloud fraction varies across different atmospheric regimes. This variation underscores the complex relationship between PV and cloud dynamics. It suggests that while PV may serve as a reliable indicator of atmospheric movement, cloud coverage responses are modulated by additional factors specific to each regime. Therefore, using the PV to determine the transition rate instead of the cloud coverage or streamfunction is a natural choice, which builds the dependence of these fields on the PV but incorporates randomness to account for the contributions from other factors. To summarize, the PV is adopted as a foundational parameter for determining transition rates in a Markov transition process.

\subsubsection{Gaussian kernel}
Within the PQG framework, the connection between PV and thermodynamic variables is established through the inversion of the elliptic equation, denoted by $\nabla^{-2}$ (check equations \eqref{eqn:pv-m-inversion-1} and \eqref{eqn:pv-m-inversion-2} in Section~\ref{ap:A}), which serves as a smoothing mechanism in the system. By inverting this elliptic equation, the process effectively distributes the influence of PV over a broader area, reducing sharp gradients and creating a more continuous representation of other variables, such as the streamfunction $\psi$ and total water $q_t$. Maintaining this smoothing mechanism is crucial for capturing large-scale, coherent structures in atmospheric dynamics, thereby providing a more accurate model of the spatiotemporal behavior of precipitation. One alternative in the stochastic model for this mechanism is applying a Gaussian kernel to process PV fields for determining transition rates, as outlined in equations~\eqref{eq:tansition-gauss-1} to~\eqref{eq:tansition-gauss-2}. Additionally, using a Gaussian kernel (blurring) facilitates detecting and analyzing anomalies or irregularities within PV fields. Such anomalies might be obscured in high-resolution data due to noise or other disturbances. This technique smooths out these disturbances, making the underlying patterns more evident.

The Gaussian kernel is imposed on the transitions rates:
\begin{align}
    \mu(x,y) = \mathcal{F}_{\mu}(G \circ PV(x,y)),\quad \nu(x,y) = \mathcal{F}_{\nu}(G \circ PV(x,y)),
    \label{eq:markov-transition-rate}
\end{align}
where $\mu$ is the transition rate from unsaturated state to saturated state, $\nu$ is the transition rate from saturated state to unsaturated state, $G$ is a Gaussian kernel \cite{babaud1986uniqueness}, and $\mathcal{F}_{\mu}$, $\mathcal{F}_{\nu}$ are two functions that map the Gaussian convolutioned quantities to the transition rates. The Gaussian kernel is employed to blur the impact of the PV fields on the transition rates. It facilitates the recovery of smoothed large-scale coherent structures and enhances the representation of thermodynamic variables in the PQG system.

\subsubsection{Maintaining the statistics of coherent structures using an adaptive procedure}
The Markov jump process characterizes the phase change transitions at each location $(x,y)$ and facilitates capturing the dynamical properties of the original PQG system. However, by design, the Markov jump process does not guarantee to maintain the statistics of coherent structures across the entire domain.

Specifically, it may not uniformly preserve the cloud fraction in the whole domain. To address this issue, an adaptive mechanism is incorporated into the transition rates:
\begin{align}
    \mathcal{X}_t(x,y) \to \mathcal{A}_t (\mathcal{X}_t(x,y)), \forall (x,y)\in\Omega \label{eq:adaptive}
\end{align}
where $\mathcal{A}_t$ represents the adaptive procedure designed to modify the transition rates within the stochastic process $\mathcal{X}_t(x,y)$. This approach ensures the maintenance of the target cloud fraction by dynamically adjusting the transition rates as necessary.

With the stochastic framework, the expensive PV-and-M inversion using the iterative method is replaced by the efficient Markov jump process. The efficiency gained through this method does not come at the expense of accuracy, offering a balanced solution to previous computational challenges. Notably, as the Markov jump process is run independently at grid points, it enables simulations to be conducted at coarser resolutions, which further reduces the computational cost.

To summarize, the SPQG model consists of three essential components with the details described in Section \ref{Sec:details}:
\begin{enumerate}
    \item Implementing a Markov jump process at each grid point and modeling transitions between saturated and unsaturated states provides a dynamic framework for understanding moisture variability.
    \item Employing a Gaussian kernel to associate PV with transition rates in the Markov jump process ensures that the representation of rainwater distribution is both precise and aligned with physical reality.
    \item Using an adaptive method to accurately identify the appropriate rain regime ensures that the model outputs remain closely tied to observable atmospheric phenomena and capture the statistics of coherent structures.
\end{enumerate}

\begin{figure*}
    \centering
    \includegraphics[width=.8\textwidth]{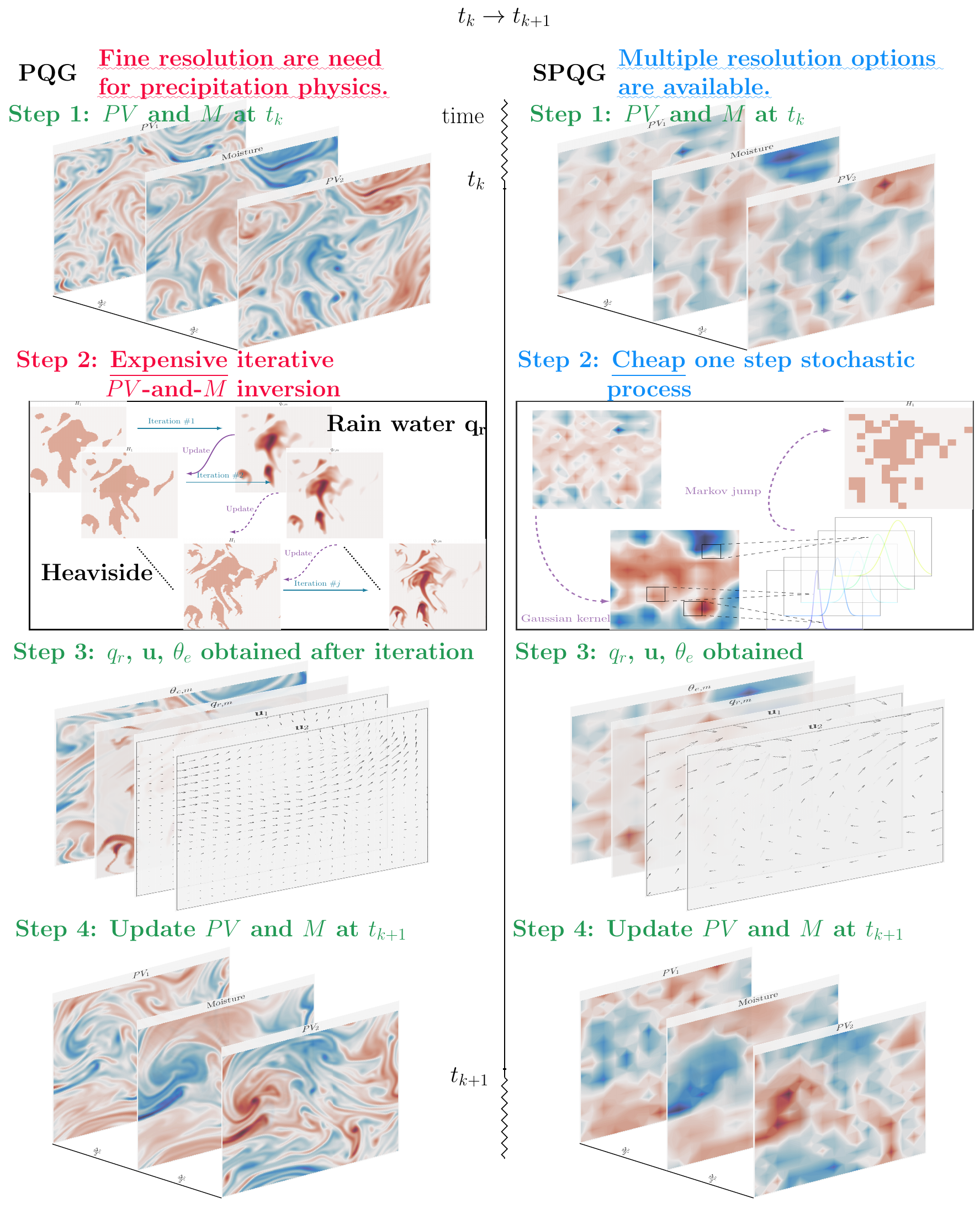}
    \caption{Schematic illustration of the PQG model (left) and the SPQG model (right).}
    \label{stochastic-chart}
\end{figure*}

A comparison between the SPQG model and the original PQG model is shown in Figure~\ref{stochastic-chart}.

\subsection{Implementation details}\label{Sec:details}

\subsubsection{Markov Jump Process: Transition in Cloud Physics}
\label{ss-markov}
A Markov jump process is a stochastic process that characterizes systems transitioning between discrete states over continuous time. Unlike traditional Markov chains, which evolve in discrete time steps, Markov jump processes allow transitions to occur at random times governed by an exponential distribution. As discussed in Section~\ref{ss-stochastic}, the Markov jump process alternates randomly between the saturated state $s_t$ and the unsaturated state $u_t$, with the subscript indicating the specific time instance $t$. The transitions are characterized by probabilities over a short time interval $\Delta t$, transitioning from time $t$ to $t + \Delta t$. The conditional probabilities for these transitions are formulated as follows:
\begin{align}
    &P(X_{t+\Delta t} = u_{t+\Delta t} | X_{t} = s_t) = \nu\Delta t +\mathrm{o} (\Delta t) ,
    \\
    &P(X_{t+\Delta t} = s_{t+\Delta t} | X_{t} = u_t) = \mu\Delta t +\mathrm{o} (\Delta t) ,
\\
    &P(X_{t+\Delta t} = s_{t+\Delta t} | X_{t} = s_t) = 1- \nu\Delta t +\mathrm{o} (\Delta t)     ,
    \\
    &P(X_{t+\Delta t} = u_{t+\Delta t} | X_{t} = u_t) = 1- \mu\Delta t +\mathrm{o} (\Delta t)     .
\end{align}
Specifically, $\mu$ represents the transition rate at a grid point from an unsaturated to a saturated state, essentially quantifying the propensity for cloud formation under given atmospheric conditions. Conversely, $\nu$ quantifies the transition rate from a saturated to an unsaturated state, reflecting the likelihood of cloud dissipation. The magnitudes of these parameters are crucial; higher values indicate a greater likelihood of state changes within the specified time interval, $\Delta t$.

Following the description of PQG model in Section~\ref{ss-pqg}, the following relations between the transition rates and the PV are required to be satisfied for the Markov jump process:
\begin{itemize}
    \item If a grid point \((x,y)\) is saturated and has a relative large positive PV value, it is highly likely to remain saturated in the subsequent time step.
    \item Conversely, if a saturated grid point \((x,y)\) has a relative small or negative PV value, there's a significant probability it will transition to an unsaturated state in the next time step.
    \item For unsaturated grid points with a large positive PV value, the likelihood of transitioning to a saturated state increases, indicating a dynamic interchange between these atmospheric conditions.
    \item An unsaturated grid point \((x,y)\) with a large positive PV value also has a high probability of remaining in its current state.
\end{itemize}

Therefore it is reasonable to assume that transition rates $\mu$ and $\nu$ are functions of the PV, which is described in equation
\eqref{eq:markov-transition-rate}. Hyperbolic tangent functions are used to describe $\mathcal{F}_{\mu}$ and $\mathcal{F}_{\nu}$ in \eqref{eq:markov-transition-rate}:
\begin{align}
      \mu(x,y) &= \mathcal{F}_{\mu}(G \circ PV(x,y)) =\tanh(a_1(G \circ PV(x,y)) +b_1)  ,\label{eq:tansition-gauss-1}
      \\
      \nu(x,y) &= \mathcal{F}_{\nu}(G \circ PV(x,y)) =\tanh(a_2(G \circ PV(x,y)) +b_2)  , \label{eq:tansition-gauss-2}
\end{align}
where $a_1,a_2,b_1,b_2$ are parameters which leverage the tuning points of transition rates from the PV and $G$ is the Gaussian kernel.

This modeling framework supposes that PV inherently provides a correlation among various grid points, obviating the need for explicit modeling of spatial correlations among random noise components. By leveraging the natural spatial coherence indicated by PV, this approach simplifies the simulation process, enhancing model efficiency without compromising the accuracy of atmospheric dynamics representation. The hyperbolic tangent function ($\tanh$) is employed to map the PV and transition rate in stochastic models, for it is smooth and differentiable across its domain, which can realistically represent gradual changes between saturated and unsaturated states. The tanh function is particularly suited for delineating different phase space environments, such as saturated regions filled with liquid water and unsaturated regions with only water vapor.

\subsubsection{Gaussian Kernels: {Coherent Structures}\label{ss-gauss}}

The Gaussian kernel is an essential tool in image processing, predominantly used for its capability to blur images \cite{babaud1986uniqueness,wand1994kernel,wilson2013gaussian}. The primary function of this kernel is to impart a smoothing effect on images, which is instrumental in minimizing noise and unnecessary details. This effect is achieved through the Gaussian function:
\begin{align}\label{Gaussian_kernel}
    G(x, y) = \frac{1}{2\pi\sigma^2} e^{-\frac{x^2 + y^2}{2\sigma^2}}
\end{align}
where $x$ and $y$ represent the distances from the origin along the horizontal and vertical axes, respectively, and $\sigma$ is the standard deviation of the Gaussian distribution. This kernel is methodically applied to every pixel in an image to generate a uniformly blurred effect, establishing it as a fundamental technique in various image-processing tasks.

\subsubsection{Adaptive Stochastic Parametrization of Cloud Fraction\label{ss-adaptive}}

By determining the transition rate as a function of the PV fields through the application of the Gaussian kernel, a precise thermodynamic relationship is established within the SPQG framework. This approach ensures that for any given point $(x,y)$, the cloud physics is intricately retained. However, when considering the simulation over the entire domain $\Omega$, adherence to the cloud coverage expected for a particular regime may not be consistently maintained.

To address this, the adaptive stochastic parametrization method, as referenced in equation \eqref{eq:adaptive}, is utilized to ensure the spatial consistency of cloud physical properties. This method parallels the principles of adaptive Markov Chain Monte Carlo (MCMC), where the proposal distribution dynamically adjusts based on the sampled data to enhance convergence and sampling efficiency. Similarly, in the SPQG framework, the adaptive stochastic parametrization method adjusts parameters in real time to maintain the desired cloud coverage and consistency of cloud physical properties across the entire domain. This dynamic adjustment is crucial for capturing the complex and evolving nature of atmospheric dynamics, ensuring that the model remains accurate and reliable.

More precisely, this adaptive method iteratively modifies the transition rates, $\mu$ and $\nu$, in Markov jump processes until the cloud fraction at time $t$ is within the range $[\mu_c - \sigma_c, \mu_c + \sigma_c]$, where $\mu_c$ and $\sigma_c$ are the mean and standard deviation of the cloud fraction from the PQG data. This yields the following form:
\begin{widetext}
\begin{equation}\label{adaptive_rate}
\mathcal{X}_t(x,y) \to \mathcal{A}_t (\mathcal{X}_t(x,y)):\quad
\begin{cases}
    \text{if cloud fraction } > \mu_c + \sigma_c,  \text{ adjust }
    \begin{cases}
        \nu = \nu - \epsilon\\
        \mu = \mu + \epsilon\\
    \end{cases} \\
    \text{if cloud fraction } < \mu_c - \sigma_c,  \text{ adjust }
    \begin{cases}
        \nu = \nu + \epsilon\\
        \mu = \mu - \epsilon\\
    \end{cases}
\end{cases}
\end{equation}
\end{widetext}
where $0 < \epsilon \ll 1/\Delta t$ is the penalty term applied to the transition rates, chosen based on the specific case.
This adaptive stochastic parametrization method is essential for fine-tuning the cloud structure, ensuring that its coherent form is maintained while adhering to the prescribed cloud regime. Within the SPQG framework, the adaptive method dynamically adjusts transition rates to sustain the desired cloud coverage and physical consistency across the entire domain. Such dynamic adjustment is crucial for accurately capturing the complex and evolving nature of atmospheric dynamics, thereby enhancing the reliability of the model.


\section{Numerical Results \label{ss-spqg-numerical}}

This section will elaborate on the numerical experiments conducted within the SPQG model framework as presented in Section~\ref{ss-spqg}. While leveraging the SPQG model to significantly improve the computational efficiency, it is also crucial to ensure the physical reliability of the model. The preservation of physical properties within the SPQG model is attributed not only to the calibration of the transition dynamics within the Markov jump process exploiting the PV values but also to its two key components: the Gaussian kernel and the adaptive stochastic parametrization method. Therefore, in addition to showing the skillful simulation from the SPQG model, the essential role of each of the three components will be demonstrated via systematic numerical experiments.

In the following, the spatial resolution of the original PQG model is $128\times 128$. As was presented above, the SPQG allows an accurate model simulation with a much lower resolution. Therefore, a $16\times 16$ spatial resolution is adopted in the SPQG model for a rapid simulation.

\subsection{Parameters Settings}
\label{ss-spqg-parameter}

The conventional setup of standardized environmental parameters (see Appendix~\ref{ap:A}) as employed in the PQG model framework under mid-latitude atmospheric conditions are given in Table~\ref{table:param-spqg}. This regime aligns with the parameters utilized in the PQG simulations as detailed by Hu et al.~\cite{hu2021initial}.

\begin{table}[H]
    \centering
    \begin{tabular}{c|c|c|c|c|c|c}
    \hline
     Parameter& $N$  &$\beta$ & $G_M$  &$L$ &$L_{ds}$&$L_{du}$ \\
      \hline
     Value &$128$ &$2.5$ & $1.0$ & $1.0$ &$1/\sqrt{2}$&$1/2$
     \\
         \hline
         \hline
     Parameter& $V_r$  &$E$ &$\Delta z$ & $\kappa$ & $U$ & $\nu$\\
      \hline
     Value &$1.0$ &$0.2$& $0.5$  & $0.05$ & $0.1$ & $8.39\times 10^{-8}$
    \\
      \hline
    \end{tabular}
    \caption{Parameter values for the numerical tests in the original PQG model.
    }
    \label{table:param-spqg}
\end{table}
The regime focuses here has an evaluation rate of $0.2$. The conclusions from the SPQG model in other parameter regimes remain qualitatively similar. Exploiting on the data from the original PQG simulation, the calibrated parameter values of the transition rate in \eqref{eq:tansition-gauss-1}--\eqref{eq:tansition-gauss-2} for the Markov jump process are $a_1=a_2= 0.2093$ and $b_1=b_2=0$. The parameters for the adaptive procedure \eqref{adaptive_rate} are $\mu_c = 25.87\%$, $\sigma_c = 0.0350$ and $\epsilon = 0.0175$. Finally, since a discrete mesh grid is utilized, the Gaussian kernel $G(x,y)$ in \eqref{Gaussian_kernel} is given by a matrix. Here, a local $3\times3$ matrix is adopted to smooth out the value in the center:
\[
G = \frac{1}{16}
\begin{bmatrix}
   1 & 2 & 1 \\
   2 & 4 & 2 \\
   1 & 2 & 1
\end{bmatrix}.
\]

\subsection{Predictability of phase boundaries in the PQG model}\label{ss-predict}
\begin{figure*}
    \centering
    \includegraphics[width=.8\textwidth]{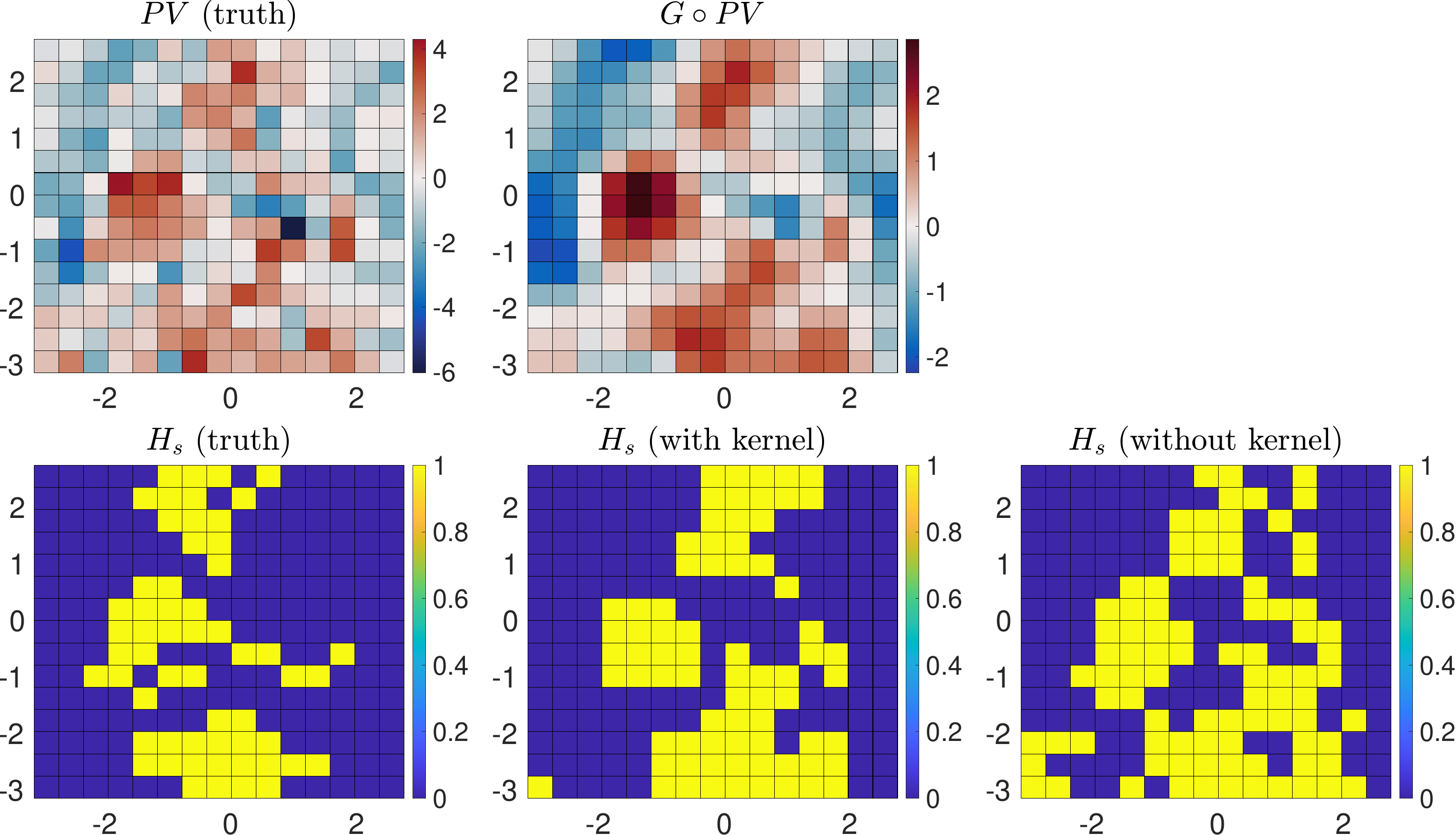}
     \caption{Comparison of the PV and Heaviside functions. Left column: one snapshot from the PQG model. Middle column: applying the Gaussian kernel to the PQG PV field and the associated Heaviside function computed from the Markov jump process using the SPQG model. Right column: the Heaviside function calculated from the SPQG model without applying the Gaussian kernel smoothing to the PV field.  }
     \label{fig:predicted-1-a}
\end{figure*}

One crucial component of the SPQG framework is the deployment of Markov jump stochastic processes to precisely simulate the phase transition of water molecules from the vapor to the liquid state. The capacity for accurately simulating the Heaviside function is pivotal to ensuring the efficacy of the stochastic model. Figure~\ref{fig:predicted-1-a} validates the transition dynamics of the Markov jump process---namely, the calibrated transition rates utilizing the PV values---for accurately predicting the Heaviside function and, subsequently, the phase interfaces that separate saturated from unsaturated regions. The first column in Figure~\ref{fig:predicted-1-a} shows one snapshot of the PV and $H_s$ from the PQG model. The second column shows the counterpart from the SQPG model. It is worth noting that the results in the second column are not from a free run of the SPQG model. Instead, it is given by plugging the true PV value from the first column to \eqref{eq:markov-transition-rate} and then simulating the $H_s$ field based on the resulting transition rates. This allows us to understand the capability of applying the Markov jump process with the true PV field in recovering the phase boundary. Although different random numbers lead to distinct results of $H_s$, all of them are qualitatively similar to the pattern shown in the second column of Figure~\ref{fig:predicted-1-a}. In comparison, in the absence of kernel smoothing, the coherent structure of $H_s$ is broken, which can be seen in the third column of this figure. The results confirm the necessity of implementing the Gaussian kernel to identify the phase interfaces.

\subsection{Numerical simulation of the SPQG model}
\label{ss-spqg-physical}

The results in this subsection are obtained by running the SPQG model forward in time. The focus is on understanding the effectiveness of the SPQG model in characterizing essential physical quantities, including PV, wind fields, total water content, and their intrinsic coupling relationships.

\begin{figure*}
    \centering
    \includegraphics[width=.8\textwidth]{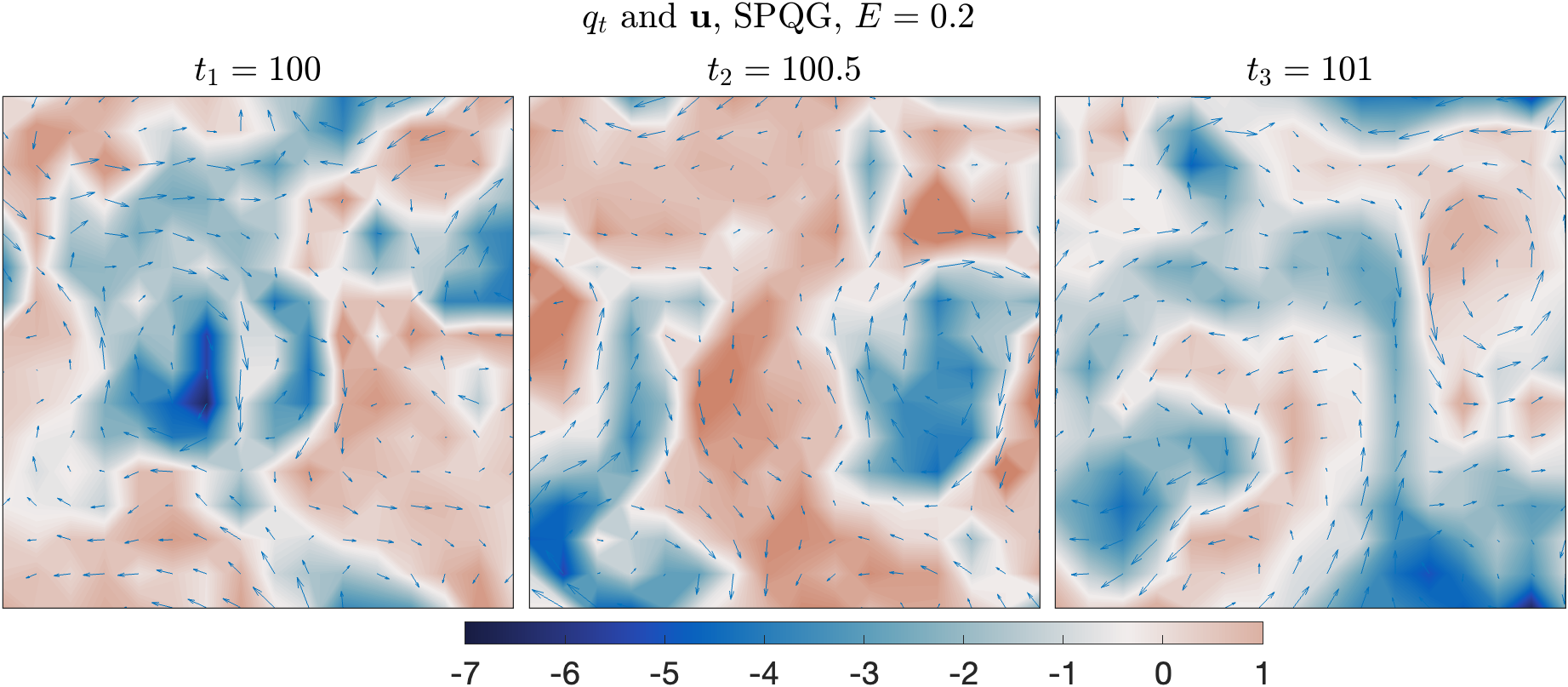}
    \includegraphics[width=.8\textwidth]{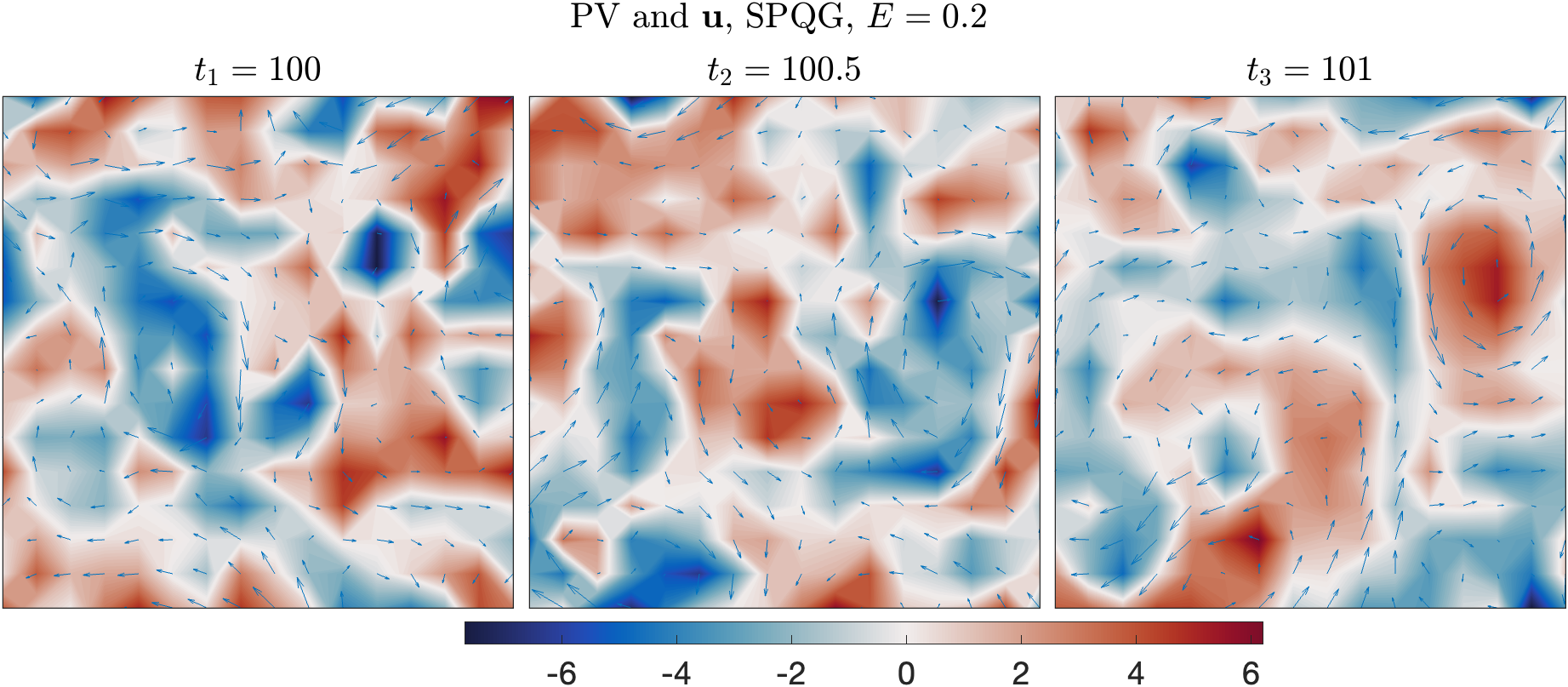}
     \caption{Results of the SPQG model: snapshots of the PV, the $q_t$ and the velocity field at different times. All spatially 2D plots are fixed-time, $(x,y)$-slices of fields at the intermediate height between levels 1 and 2, where $x$ $(y)$ is the zonal (meridional) direction.
     }
     \label{fig:spqg-evap-2-a}
\end{figure*}

\begin{figure*}
    \centering
    \includegraphics[width=.8\textwidth]{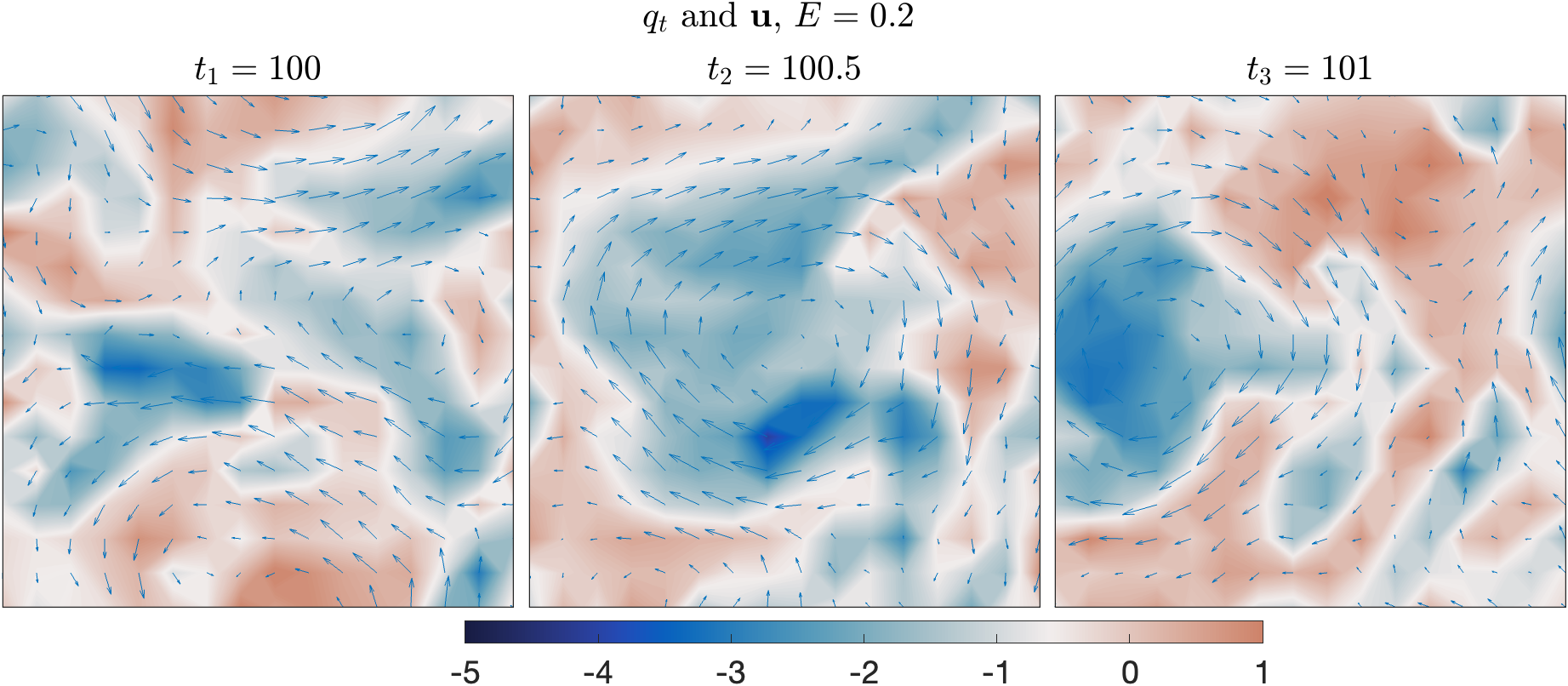}
    \includegraphics[width=.8\textwidth]{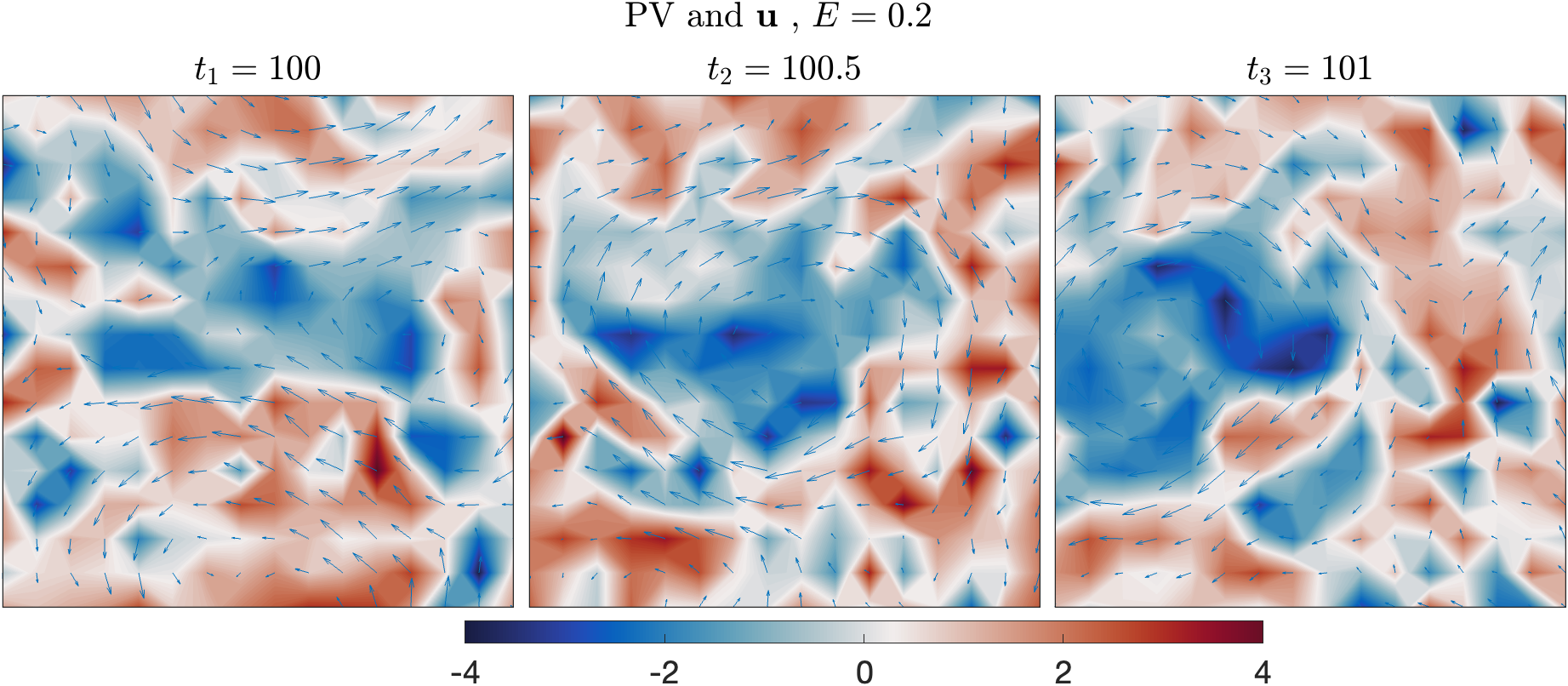}
     \caption{Results of the PQG model: snapshots of the PV, the $q_t$ and velocity field at different times.
     }
     \label{fig:pqg-evap-2-a}
\end{figure*}

Figure~\ref{fig:spqg-evap-2-a} exhibits the distributions of the PV and the total water content \(q_t\) set on top of the velocity field $\bf u$ at various time intervals from the SPQG simulation.
Conversely, Figure~\ref{fig:pqg-evap-2-a} portrays the same variables from the original PQG model. From these two sets of simulations, the coupling relationships between the represented physical quantities bear striking resemblance. The patterns of the total water content and the velocity field are consistent in both models. In Figure~\ref{fig:spqg-evap-2-a}, regions with relatively high/low \(q_t\) values indicate a tendency toward cyclonic/anticyclonic motions in the velocity field. This observation aligns with insights derived from the high-resolution PQG data in Figure~\ref{fig:pqg-evap-2-a}. Additionally, the PV in both frameworks shows similar relationships with the velocity field, where regions with relatively high/low PV values tend toward cyclonic/anticyclonic motions.

\begin{figure*}
    \centering
    \includegraphics[width=.35\textwidth]{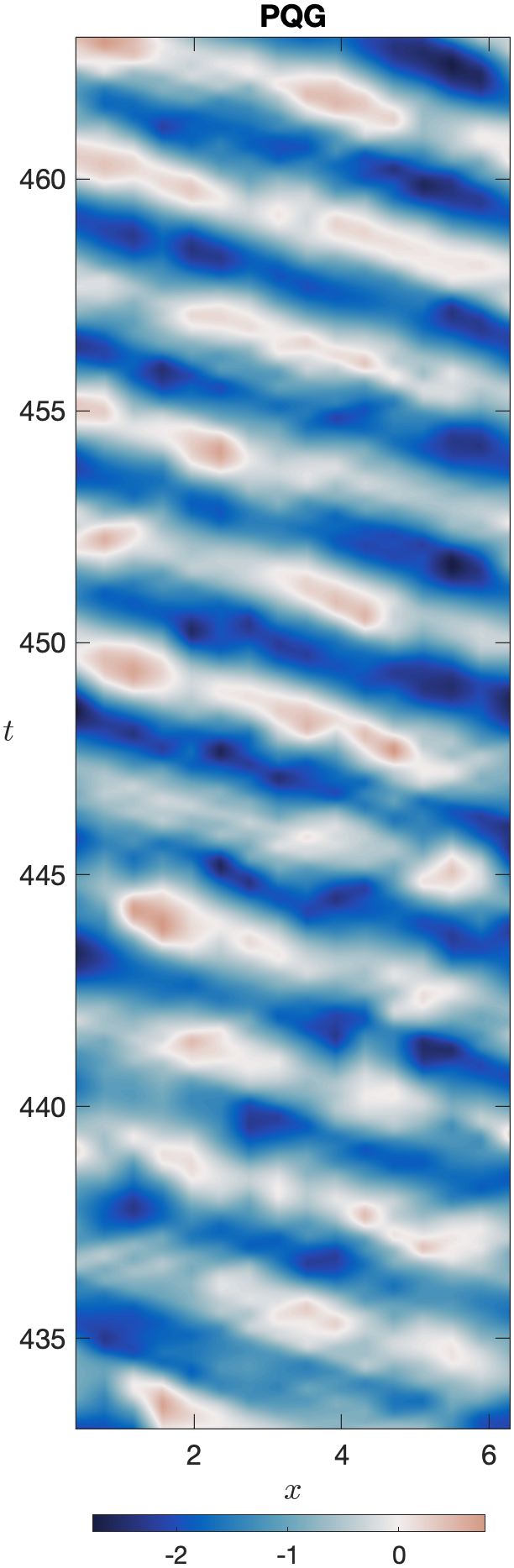}
    \includegraphics[width=.35\textwidth]{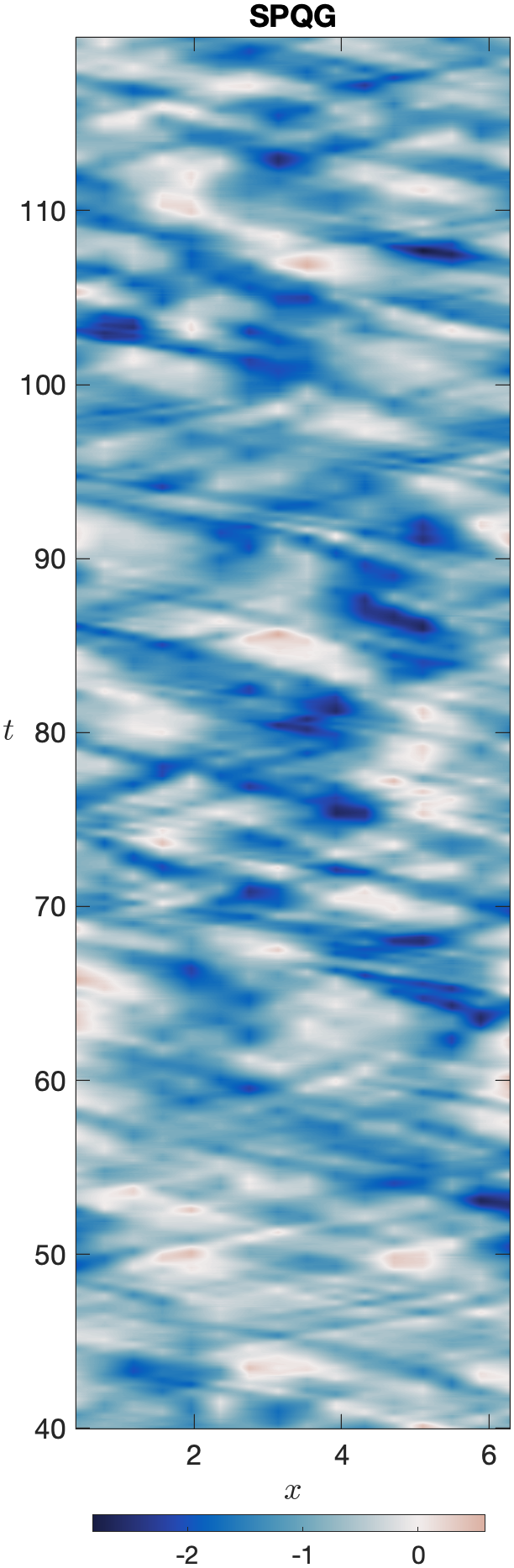}
     \caption{Meridionally-averaged $\bar{q}_t(t,x)$ field for the PQG model (left) and the SPQG model (right).
     }
     \label{fig:spqg-evap-2-d}
\end{figure*}

Next, one crucial dynamical property of the PQG model is its west-to-east pattern of cloud movement. This can be seen from the left panel of Figure~\ref{fig:spqg-evap-2-d}, which shows the meridionally-averaged \(\bar{q}_t(t,x)\) field. Despite slightly more chaotic patterns than the PQG simulation due to the stochastic noise, The SPQG model captures such an important feature. In particular, the propagation angle and speed of the clouds are relatively similar to those from the PQG model. See the right panel of Figure~\ref{fig:spqg-evap-2-d}. It is worth noting that the overall features in the PQG model (propagation angle and speed) are determined by environmental background configurations, such as the background profile of vertical shear of flow in the zonal direction and the background moisture profile. These factors, among the core mechanisms that drive the entire turbulent system, are not directly affected by the stochastic noises. Therefore, the SPQG model, by design, reproduces the overall structure and core dynamic mechanisms.

\begin{figure*}
    \centering
    \includegraphics[width=.8\textwidth]{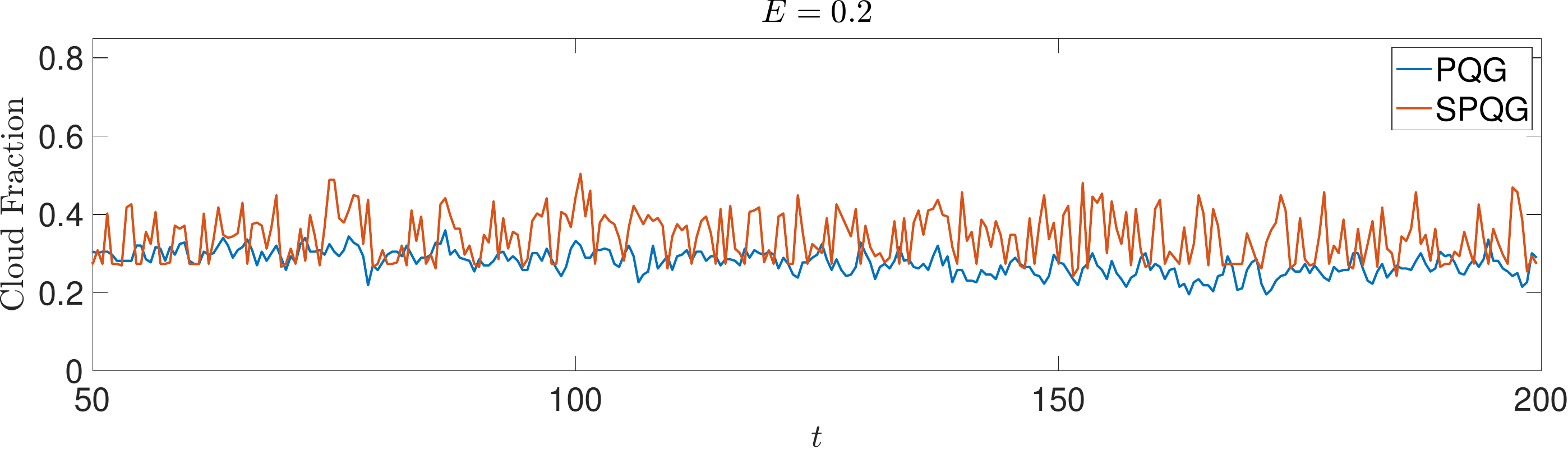}
     \caption{Time series of cloud fraction for the PQG model (blue curve) and the SPQG model (red curve).}
     \label{fig:spqg-evap-2-c}
\end{figure*}

Figure~\ref{fig:spqg-evap-2-c} shows the time evolution of the cloud fraction in the two models. The result reveals that the SPQG model captures the cloud fraction dynamics, including both the mean and the temporal fluctuation profile, thanks to the adaptive mechanism in the stochastic PQG model.

Finally, the gross features of the PQG and SPQG jets are compared in the top panel of Figure~\ref{fig:spqg-evap-2-c} and the top panel of Figure~\ref{fig:pqg-evap-3-a}. Qualitatively, the SPQG model captures the PQG structure consisting of two jets, one dominant easterly (westward propagating) jet and one dominant westerly (eastward propagating) jet.  As representative statistics, we quantitatively compare the proportion of the domain occupied by the jets (Table~\ref{tab:width-jet}), as well as their temporal means and standard deviations (Table~\ref{tab:mag-jet}). Table~\ref{tab:width-jet} shows $98\%$ agreement for the mean widths and more than $43\%$ agreement for the standard deviation of the proportion of the domain occupied by both westerly and easterly jets, while Table~\ref{tab:mag-jet} shows better than $88\%$ for means and better than $44\%$ for standard deviations.  This agreement is encouraging, given the simplicity of the SPQG model, and suggests that further improvements in the model could lead to highly accurate results.  For example, the Gaussian kernel can be further improved using a machine learning algorithm.

\subsection{The role of the key components in the SPQG model}
\label{ss-spqg-component}

Recall the three components in the SPQG model: (a) using the Markov jump process to compute the Heaviside function, (b) applying the Gaussian kernel to obtain realistic large-scale spatial patterns and phase boundaries, and (c) incorporating the adaptive mechanism to guarantee the model capturing crucial statistics, such as the cloud fraction. The vital role of the transition dynamics driven by the Markov jump process in reducing the computational cost has been revealed in the previous section. In this section, our analysis will be extended to study the two other key components.

The PQG model is adept at depicting the behaviors of mid-latitude jet variability, such as the poleward propagation of the latitude of the jet. In light of this core atmospheric physical characteristic as a standard of examination, we conduct numerical simulations to demonstrate the essential capabilities of the following three model configurations in describing jet stream dynamics:
\begin{enumerate}
    \item \textbf{SPQG:} The SPQG model incorporating all the key components.
    \item \textbf{SPQG-NG:} A simplified SPQG model without the Gaussian kernel.
    \item \textbf{SPQG-NA:} A simplified SPQG model without the adaptive mechanism for the cloud fraction adjustments.
\end{enumerate}

\begin{figure*}
    \centering
    \includegraphics[width=.9\textwidth]{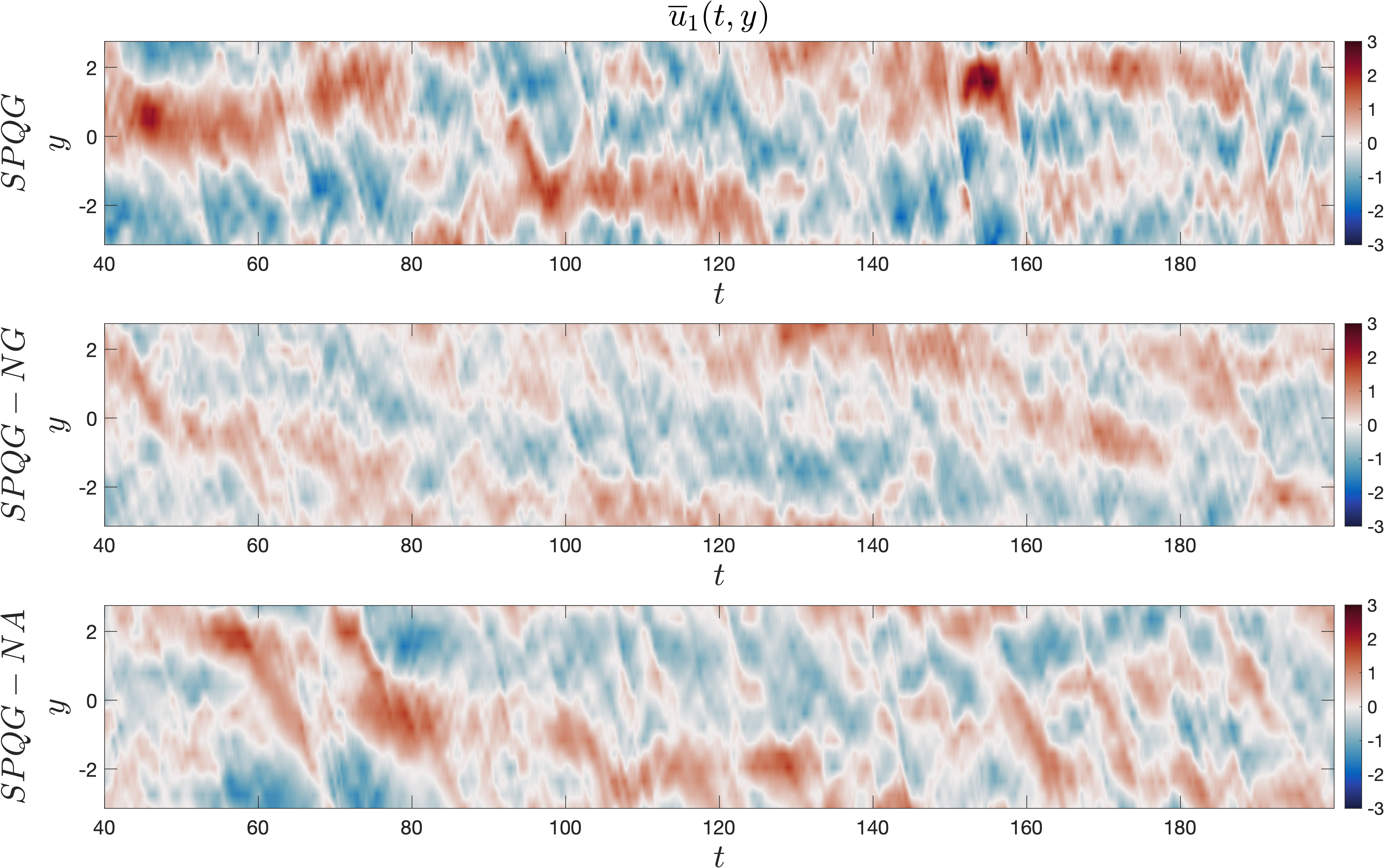}
     \caption{
     Zonal mean velocity field simulated by the three different stochastic models. All simulations are executed with the same random number seed to guarantee a fair comparison.
     }
     \label{fig:spqg-20-jet}
\end{figure*}

Figure~\ref{fig:spqg-20-jet} compares the zonal mean jet profiles using the three stochastic models. All simulations are executed with the same random number seed to guarantee a fair comparison. This consistency allows for a controlled evaluation of the influence of each separate component on the model performance. The top panel of Figure~\ref{fig:spqg-20-jet}, which shows the results from the SPQG model, is regarded as a benchmark. In the absence of the Gaussian kernel (SPQG-NG) and adaptive mechanism (SPQG-NA), the large-scale coherent structure of the zonal jet is disrupted over time. Specifically, a more fragmented jet stream with numerous small-scale features is seen.
Thus, both the Gaussian kernel in smoothing the PV field for determining the transition rates and the adaptive adjustment of cloud fraction have substantial impacts on preserving the coherent structure and the dynamic movement of the jet stream. Furthermore, ignoring the Gaussian kernel in the SPQG-NG model affects the coherent structure and the intensity of the zonal jet. Ignoring the Gaussian kernel leads to significant deviations in jet magnitudes compared to both PQG and the complete SPQG model, indirectly reflecting the key role of the kernel component in accurately capturing the phase transition dynamics.

{
\color{blue}
\begin{table}[H]
    \centering
    \begin{tabular}{c|c|c|c|c}
    \hline\hline
           &  \multicolumn{2}{c|}{Westerly} &\multicolumn{2}{c}{Easterly}\\ \cline{2-5} Model & mean   & std & mean   & std \\   \hline 
        PQG &5.0e-1 &5.4e-2  &5.0e-1 & 5.5e-2\\
        SPQG& 4.9e-1& 8.5e-2 &5.1e-1 &8.5e-2\\\hline
    \end{tabular}
    \caption{Mean and standard deviation (std) for temporal variation of proportion of the domain for westerly and easterly jets in PQG and SPQG.
The proportion of the domain for the westerly (easterly) jet is defined as the ratio of the length of positive (negative) zonal wind to the total width of the meridional side of the domain.    
}
    \label{tab:width-jet}
\end{table}

\begin{table}[H]
    \centering
    \begin{tabular}{c|c|c|c|c}
    \hline\hline
           &  \multicolumn{2}{c|}{Westerly} &\multicolumn{2}{c}{Easterly}\\ \cline{2-5} Model & mean   & std & mean   & std \\   \hline 
        PQG &1.3e0 & 3.0e-1 &-1.4e0 & 3.6e-1\\
        SPQG& 1.1e0 &4.7e-1 &-1.1e0 &3.9e-1 \\ \hline 
    \end{tabular}
    \caption{Mean and standard deviation (std) for temporal variation of maximum velocities of westerly and easterly jets for PQG and SPQG.}
    \label{tab:mag-jet}
\end{table}

}

\section{Conclusion\label{ss-conclusion}}
In this paper, an SPQG system is developed that captures crucial features of the PQG dynamics and significantly reduces the computational cost by avoiding using the expensive iterative solver in the PQG system. Exploiting the Markov jump process to represent transitions between saturated and unsaturated states at individual grid points, guided by the PV field processed through a Gaussian kernel, establishes a robust method for capturing the spatial variability and temporal dynamics of cloud formation and dissipation. This approach not only aligns with physical observations but also enhances the capacity of the model to represent complex thermodynamic relationships. Moreover, employing an adaptive mechanism to maintain the spatial consistency of cloud physical properties across the entire domain further allows the model to maintain the consistent spatial pattern of precipitation and reproduce large-scale statistical features.

The numerical tests demonstrate that the SPQG model captures the essential characteristics of the original PQG framework with a much lower computational cost. Notably, the SPQG model reproduces the precipitation patterns as in the PQG model, especially in accurately mirroring both the spatial distribution and the temporal variability of rainfall. Furthermore, the tests confirm that the SPQG can simulate key dynamic features akin to those of the original QG model, including the PV fields, jet streams, and zonal mean flows, highlighting its effectiveness in simulating atmospheric dynamics with computational efficiency.

There are several future directions that may provide potential improvement of the current framework. First, expanding the stochastic framework to incorporate the moist Boussinesq model presents an opportunity to delve deeper into the dynamics of moist atmospheric processes \cite{zhang2021fast, zhang2022convergence, zhang2021effects, hernandez2014investigation, deng2012tropical}. This extension would allow for a more detailed understanding of the interaction between moisture, temperature, and atmospheric motions, offering a comprehensive approach to modeling atmospheric convection and cloud formation within a theoretically robust framework \cite{yau1996short, liu2023parameterization}. Second, utilizing the SPQG as a surrogate forecast model for precipitation systems can lead to more effective and efficient data assimilation schemes~\cite{mou2023combining, chen2014predicting,chen2023stochastic}, especially in capturing the precipitation patterns due to the complex thermodynamics. This capability is also potentially useful in addressing and mitigating the impacts of extreme weather events \cite{lopez2002implementation,gottschalck2005analysis}.

\section*{Acknowledgements} N.C. is grateful to acknowledge the support of the Army Research Office (ARO) W911NF-23-1-0118. 
Y.Z is grateful to acknowledge the support of National Natural Science Foundation of China grants 12241103 and Shanghai Pujiang Program grant 22PJ1403500. L.M.S received support from the National Science Foundation, Division of Mathematical Sciences DMS-1907667, as well as Deutsche Forschungsgemeinschaft (DFG) through the Research Unit FOR5528.

\section*{Data Availability}
The data that support the findings of this study are available from the corresponding author upon reasonable request.

\appendix
\section{Appendix A.}
\label{ap:A}

In this appendix, we present the detailed configuration of the PQG model constrained to two vertical levels. Additional features have been incorporated into the dynamical model in \eqref{eqn:PQG-layer-1}--\eqref{eqn:PQG-layer-3}, such as an evaporation source term, lower-level friction, and background velocity and temperature to provide conditions for baroclinic instability. This setup is similar to other studies of two-level QG equations~\cite{edwards2020spectra,qi2016low}, although the present model also includes phase changes. Similar to the approach used for the dry variant, the PQG system is obtained by applying a staggered vertical grid, as detailed in \cite{hu2021initial}. In the two-level model, the subscript \((\cdot)_{j}\), where \(j=1,2\), indicates variables assigned to either the first or second level, while the subscript \((\cdot)_{m}\) denotes variables situated at the intermediate level. The governing equations are given by:
\begin{widetext}
\begin{align}
    &\begin{aligned}
     \frac{\partial  { PV_{1}}}{\partial t}
     +
     J(\psi_1,{ PV_{1}})
   -
   U\frac{\partial { PV_{1}}}{\partial x}
   +
   &\beta v_1
   +
      v_1\frac{\partial { PV_{1,bg}}}{\partial y}
    =
    \\
     -
    &{
    \frac{L_{du}}{L_{ds}}\frac{L}{L_{ds}}\frac{\partial\textbf{u}_{h,1}}{\partial z}\cdot\nabla_h\theta_{e,1}
    }
    -\kappa\Delta_h \psi_1
   -\nu\Delta_h^{4} {PV_{1}},
\end{aligned}
    \label{eqn:layer-pqg-1}
    \\
    &\begin{aligned}
     \frac{\partial { PV_{2}}}{\partial t}
     +
   J(\psi_2, {PV_{2}})
    +
    U\frac{\partial  { PV_{2}}}{\partial x}
    +
    &\beta v_2
    +      v_2\frac{\partial { PV_{2,bg}}}{\partial y}
    =
    \\
    -
    &{
    \frac{L_{du}}{L_{ds}}\frac{L}{L_{ds}}\frac{\partial\textbf{u}_{h,2}}{\partial z}\cdot\nabla_h\theta_{e,2}
    }
    -\nu\Delta_h^{4}  { PV_{2}},
\end{aligned}
    \label{eqn:layer-pqg-2}
\\
    &\begin{aligned}
    {
     \frac{D_m M_m}{Dt} +
     v_m\frac{\partial M_{bg}}{\partial y}
    =-\frac{V_r}{\Delta z}q_{r,m}-\nu\Delta_h^{4} M_m +E
    },
    \label{eqn:layer-pqg-3}
    \end{aligned}
\end{align}
\end{widetext}
where the PV variables $PV_{1,2}$ and the moisture variable $M_m$ are related to the streamfunctions $\psi_{1,2}$ by nonlinear, elliptic operators (see (\ref{eqn:pv-m-inversion-1}) and (\ref{eqn:pv-m-inversion-2}) below).  From the streamfunction $\psi_{j}$, the horizontal winds ${\bf u}_{h,j} = (u_j,v_j)$ are obtained from the relations
$u_j = -\partial \psi_j/\partial y$, $v_j = \partial \psi_j/\partial x$ with mid-level values $u_m = (u_1+u_2)/\Delta z$, $v_m = (v_1+v_2)/\Delta z$, where $\nabla_h$ is the horizontal part of the gradient operator and $\Delta z$ is the distance between levels 1 and 2.
Besides, $J(A, B)= A_x B_y - A_y B_x$ represents the Jacobian operator. 
The equivalent potential temperature $\theta_e$ (a linear combination of potential temperature and water vapor) is found from the streamfunction $\psi_j$ and $M_m$ using the relation
\begin{widetext}
\begin{equation}
\theta_{e,m} = H_s\left(\frac{L}{L_{du}}\frac{\psi_2-\psi_1}{\Delta z}+q_{vs,m}\right)+H_u\left(\frac{1}{1+G_M}\frac{L}{L_{du}}\frac{\psi_2-\psi_1}{\Delta z}+\frac{1}{1+G_M}M_m\right),
\label{eqn:pv-m-inversion-3}
\end{equation}
\end{widetext}
where $G_M$ is a non-dimensional parameter of order one ($O(1)$), associated with the background state profiles of equivalent potential temperature and total water. $L$ is the reference length scale (1000 km), and $L_{du}$ corresponds to the Rossby radius of deformation for the unsaturated background state, defined as $L_{du} = N_u H / f$. Relating this to $Ro$ and $Fr_u$, the relationship is $L_{du} = L Ro / Fr_u$.

The total water mixing ratio is $q_{t,m} = q_{v,m} + q_{r,m}$, where $q_{v,m}, q_{r,m}$ are the vapor and liquid components, respectively.
In terms of $M_m$ and $\theta_{e,m}$, $q_{t,m} = M_m - G_M \theta_{e,m}$.
A saturation profile $q_{vs,m}$ separates unsaturated regions (water vapor only) from saturated regions (vapor and rain water).  In the current two-level model, $q_{vs,m}$ may be written in  terms of the streamfunction as
\begin{align}
&q_{vs,m} = q_{vs}^0+q_{vs}^1 \frac{\psi_2-\psi_1}{\Delta z },
\label{def:qvs-2level}
\end{align}
where the parameters $q_{vs}^0$ and $q_{vs}^1$ are non-negative constants. The absence or presence of rain $q_{r,m}$ is diagnosed from $M_m, \theta_{e,m}$ and $q_{vs,m}$ according to the relation
$q_{r,m} = \max(0,M-G_M \theta_{e,m}-q_{vs,m})$.
The cloud indicator (Heaviside) functions are then
\begin{align}
    &H_s =\begin{cases}
        1 & \text{if } q_r>0\\
        0 & \text{if } q_r=0
    \end{cases},  &
        &H_u =\begin{cases}
        0 & \text{if } q_r>0\\
        1 & \text{if } q_r=0
    \end{cases},
    \label{def:cloudindicator}
\end{align}
where  $H_s=1$ ($H_s=0$) signifies that the area at coordinates $(x,y)$ is in a saturated (unsaturated) state .
At the top and bottom, $PV_{1,2}, M_m$ and all associated unknowns obey periodic boundary conditions in the horizontal directions $(x,y)$. The top and bottom boundary conditions are established based on a rigid-lid assumption (precluding flow across the boundaries), the material conservation of equivalent potential temperature $\theta_e$, and the transport equation governing total water content $q_{t,m}$, which incorporates a term accounting for rainfall.
 A simplified version replaces the latter two conditions with $q_{t,m} = \theta_{e,m} = 0$.

As alluded to above, the model assumes a prescribed background state with characteristic length-scale $L = 1000$ km. This environmental configuration is partially determinated by the parameter $G_M$, representing the ratio of vertical gradients in equivalent potential temperature and total water profiles (both linear).
In addition, $\beta$ characterizes the meridional variation of the Coriolis parameter in a narrow mid-latitude band, $(U,-U)$ is the vertical shear of flow in the zonal direction, and the length scales $L_{du}$, $L_{ds}$ are the Rossby deformation radii associated with unsaturated and saturated flow regions, which are defined as $L_{du} = N_u H / f$, $L_{ds} = N_s H / f$ respectively. Given the parameters $U$, $G_M$, $L$, $L_{du}$, and $q_{vs}^1$, the background PV profile is defined as $PV_{j,bg} = (-1)^j (1 + q_{vs}^1) \frac{1}{(\Delta z)^2} \frac{L^2}{L_{du}^2} (2Uy)$, and the background moisture profile is $M_{bg} = - (q_{vs}^1 + G_M (1 + q_{vs}^1)) \frac{1}{\Delta z} \frac{L}{L_{du}} (2Uy)$.

Specifically, $q_{vs}^1$ is a threshold parameter (\ref{def:qvs-2level}) for phase changes.
The remaining (constant) parameters in (\ref{eqn:layer-pqg-1})-(\ref{eqn:layer-pqg-3}) are the evaporation source term $E$ which regulates the cloud fraction in this setup, as well as dissipation terms parameterized by $\kappa$ (friction at the bottom level) and $\nu$ (dissipation of small-scale turbulence).

To close the PQG system, it is necessary to perform the PV-and-M inversion to find the streamfunction $\psi$ and associated quantities ${\bf u}_h, \theta_e, q_t$, etc.  In the two-level setup, the nonlinear elliptic equations relating PV and $M$ to the streamfunction $\psi$ are given by
\begin{widetext}
\begin{align}
&\begin{aligned}
    PV_{1} &= \nabla_h^2 \psi_1+
    H_s\left(
    \left(\frac{L}{L_{ds}}\frac{1}{\Delta z}\right)^2(\psi_2-\psi_1)+\frac{L_{du}}{L_{ds}}\frac{L}{L_{ds}}\frac{1}{\Delta z} q_{vs,m}
    \right)+
    \\
    &H_u\left(
    \left(\frac{L}{L_{du}}\frac{1}{\Delta z}\right)^2(\psi_2-\psi_1)+\frac{L}{L_{du}}\frac{1}{\Delta z} M_{m}
    \right)
    \label{eqn:pv-m-inversion-1}
\end{aligned}
\\
&\begin{aligned}
    PV_{2} &= \nabla_h^2 \psi_2+
    H_s\left(
    \left(\frac{L}{L_{ds}}\frac{1}{\Delta z}\right)^2(\psi_1-\psi_2)-\frac{L_{du}}{L_{ds}}\frac{L}{L_{ds}}\frac{1}{\Delta z} q_{vs,m}
    \right)+
    \\
    &H_u\left(
    \left(\frac{L}{L_{du}}\frac{1}{\Delta z}\right)^2(\psi_1-\psi_2)-\frac{L}{L_{du}}\frac{1}{\Delta z} M_{m}
    \right).
    \label{eqn:pv-m-inversion-2}
\end{aligned}
\end{align}
\end{widetext}


\bibliography{reference}

\begin{thebibliography}{67}%
\makeatletter
\providecommand \@ifxundefined [1]{%
 \@ifx{#1\undefined}
}%
\providecommand \@ifnum [1]{%
 \ifnum #1\expandafter \@firstoftwo
 \else \expandafter \@secondoftwo
 \fi
}%
\providecommand \@ifx [1]{%
 \ifx #1\expandafter \@firstoftwo
 \else \expandafter \@secondoftwo
 \fi
}%
\providecommand \natexlab [1]{#1}%
\providecommand \enquote  [1]{``#1''}%
\providecommand \bibnamefont  [1]{#1}%
\providecommand \bibfnamefont [1]{#1}%
\providecommand \citenamefont [1]{#1}%
\providecommand \href@noop [0]{\@secondoftwo}%
\providecommand \href [0]{\begingroup \@sanitize@url \@href}%
\providecommand \@href[1]{\@@startlink{#1}\@@href}%
\providecommand \@@href[1]{\endgroup#1\@@endlink}%
\providecommand \@sanitize@url [0]{\catcode `\\12\catcode `\$12\catcode
  `\&12\catcode `\#12\catcode `\^12\catcode `\_12\catcode `\%12\relax}%
\providecommand \@@startlink[1]{}%
\providecommand \@@endlink[0]{}%
\providecommand \url  [0]{\begingroup\@sanitize@url \@url }%
\providecommand \@url [1]{\endgroup\@href {#1}{\urlprefix }}%
\providecommand \urlprefix  [0]{URL }%
\providecommand \Eprint [0]{\href }%
\providecommand \doibase [0]{http://dx.doi.org/}%
\providecommand \selectlanguage [0]{\@gobble}%
\providecommand \bibinfo  [0]{\@secondoftwo}%
\providecommand \bibfield  [0]{\@secondoftwo}%
\providecommand \translation [1]{[#1]}%
\providecommand \BibitemOpen [0]{}%
\providecommand \bibitemStop [0]{}%
\providecommand \bibitemNoStop [0]{.\EOS\space}%
\providecommand \EOS [0]{\spacefactor3000\relax}%
\providecommand \BibitemShut  [1]{\csname bibitem#1\endcsname}%
\let\auto@bib@innerbib\@empty
\bibitem [{\citenamefont {Winters}\ and\ \citenamefont
  {Martin}(2014)}]{winters2014role}%
  \BibitemOpen
  \bibfield  {author} {\bibinfo {author} {\bibfnamefont {A.~C.}\ \bibnamefont
  {Winters}}\ and\ \bibinfo {author} {\bibfnamefont {J.~E.}\ \bibnamefont
  {Martin}},\ }\bibfield  {title} {\enquote {\bibinfo {title} {The role of a
  polar/subtropical jet superposition in the may 2010 nashville flood},}\
  }\href@noop {} {\bibfield  {journal} {\bibinfo  {journal} {Weather and
  forecasting}\ }\textbf {\bibinfo {volume} {29}},\ \bibinfo {pages} {954--974}
  (\bibinfo {year} {2014})}\BibitemShut {NoStop}%
\bibitem [{\citenamefont {Houze}\ \emph {et~al.}(2017)\citenamefont {Houze},
  \citenamefont {McMurdie}, \citenamefont {Rasmussen}, \citenamefont {Kumar},\
  and\ \citenamefont {Chaplin}}]{houze2017multiscale}%
  \BibitemOpen
  \bibfield  {author} {\bibinfo {author} {\bibfnamefont {R.}~\bibnamefont
  {Houze}}, \bibinfo {author} {\bibfnamefont {L.}~\bibnamefont {McMurdie}},
  \bibinfo {author} {\bibfnamefont {K.}~\bibnamefont {Rasmussen}}, \bibinfo
  {author} {\bibfnamefont {A.}~\bibnamefont {Kumar}}, \ and\ \bibinfo {author}
  {\bibfnamefont {M.}~\bibnamefont {Chaplin}},\ }\bibfield  {title} {\enquote
  {\bibinfo {title} {Multiscale aspects of the storm producing the june 2013
  flooding in uttarakhand, india},}\ }\href@noop {} {\bibfield  {journal}
  {\bibinfo  {journal} {Monthly Weather Review}\ }\textbf {\bibinfo {volume}
  {145}},\ \bibinfo {pages} {4447--4466} (\bibinfo {year} {2017})}\BibitemShut
  {NoStop}%
\bibitem [{\citenamefont {Sapsis}(2021)}]{sapsis2021statistics}%
  \BibitemOpen
  \bibfield  {author} {\bibinfo {author} {\bibfnamefont {T.~P.}\ \bibnamefont
  {Sapsis}},\ }\bibfield  {title} {\enquote {\bibinfo {title} {Statistics of
  extreme events in fluid flows and waves},}\ }\href@noop {} {\bibfield
  {journal} {\bibinfo  {journal} {Annual Review of Fluid Mechanics}\ }\textbf
  {\bibinfo {volume} {53}},\ \bibinfo {pages} {85--111} (\bibinfo {year}
  {2021})}\BibitemShut {NoStop}%
\bibitem [{\citenamefont {Vitart}\ and\ \citenamefont
  {Robertson}(2018)}]{vitart2018sub}%
  \BibitemOpen
  \bibfield  {author} {\bibinfo {author} {\bibfnamefont {F.}~\bibnamefont
  {Vitart}}\ and\ \bibinfo {author} {\bibfnamefont {A.~W.}\ \bibnamefont
  {Robertson}},\ }\bibfield  {title} {\enquote {\bibinfo {title} {The
  sub-seasonal to seasonal prediction project {(S2S)} and the prediction of
  extreme events},}\ }\href@noop {} {\bibfield  {journal} {\bibinfo  {journal}
  {npj climate and atmospheric science}\ }\textbf {\bibinfo {volume} {1}},\
  \bibinfo {pages} {3} (\bibinfo {year} {2018})}\BibitemShut {NoStop}%
\bibitem [{\citenamefont {Santoso}, \citenamefont {Mcphaden},\ and\
  \citenamefont {Cai}(2017)}]{santoso2017defining}%
  \BibitemOpen
  \bibfield  {author} {\bibinfo {author} {\bibfnamefont {A.}~\bibnamefont
  {Santoso}}, \bibinfo {author} {\bibfnamefont {M.~J.}\ \bibnamefont
  {Mcphaden}}, \ and\ \bibinfo {author} {\bibfnamefont {W.}~\bibnamefont
  {Cai}},\ }\bibfield  {title} {\enquote {\bibinfo {title} {The defining
  characteristics of {ENSO} extremes and the strong 2015/2016 {E}l
  {N}i{\~n}o},}\ }\href@noop {} {\bibfield  {journal} {\bibinfo  {journal}
  {Reviews of Geophysics}\ }\textbf {\bibinfo {volume} {55}},\ \bibinfo {pages}
  {1079--1129} (\bibinfo {year} {2017})}\BibitemShut {NoStop}%
\bibitem [{\citenamefont {Chekroun}\ \emph {et~al.}(2024)\citenamefont
  {Chekroun}, \citenamefont {Liu}, \citenamefont {Srinivasan},\ and\
  \citenamefont {McWilliams}}]{chekroun2024high}%
  \BibitemOpen
  \bibfield  {author} {\bibinfo {author} {\bibfnamefont {M.~D.}\ \bibnamefont
  {Chekroun}}, \bibinfo {author} {\bibfnamefont {H.}~\bibnamefont {Liu}},
  \bibinfo {author} {\bibfnamefont {K.}~\bibnamefont {Srinivasan}}, \ and\
  \bibinfo {author} {\bibfnamefont {J.~C.}\ \bibnamefont {McWilliams}},\
  }\bibfield  {title} {\enquote {\bibinfo {title} {The high-frequency and rare
  events barriers to neural closures of atmospheric dynamics},}\ }\href@noop {}
  {\bibfield  {journal} {\bibinfo  {journal} {Journal of Physics: Complexity}\
  }\textbf {\bibinfo {volume} {5}},\ \bibinfo {pages} {025004} (\bibinfo {year}
  {2024})}\BibitemShut {NoStop}%
\bibitem [{\citenamefont {Lucarini}\ \emph {et~al.}(2016)\citenamefont
  {Lucarini}, \citenamefont {Faranda}, \citenamefont {de~Freitas},
  \citenamefont {Holland}, \citenamefont {Kuna}, \citenamefont {Nicol},
  \citenamefont {Todd}, \citenamefont {Vaienti} \emph
  {et~al.}}]{lucarini2016extremes}%
  \BibitemOpen
  \bibfield  {author} {\bibinfo {author} {\bibfnamefont {V.}~\bibnamefont
  {Lucarini}}, \bibinfo {author} {\bibfnamefont {D.}~\bibnamefont {Faranda}},
  \bibinfo {author} {\bibfnamefont {J.~M.~M.}\ \bibnamefont {de~Freitas}},
  \bibinfo {author} {\bibfnamefont {M.}~\bibnamefont {Holland}}, \bibinfo
  {author} {\bibfnamefont {T.}~\bibnamefont {Kuna}}, \bibinfo {author}
  {\bibfnamefont {M.}~\bibnamefont {Nicol}}, \bibinfo {author} {\bibfnamefont
  {M.}~\bibnamefont {Todd}}, \bibinfo {author} {\bibfnamefont {S.}~\bibnamefont
  {Vaienti}},  \emph {et~al.},\ }\href@noop {} {\emph {\bibinfo {title}
  {Extremes and recurrence in dynamical systems}}}\ (\bibinfo  {publisher}
  {John Wiley \& Sons},\ \bibinfo {year} {2016})\BibitemShut {NoStop}%
\bibitem [{\citenamefont {Newman}\ \emph {et~al.}(2012)\citenamefont {Newman},
  \citenamefont {Kiladis}, \citenamefont {Weickmann}, \citenamefont {Ralph},\
  and\ \citenamefont {Sardeshmukh}}]{newman2012relative}%
  \BibitemOpen
  \bibfield  {author} {\bibinfo {author} {\bibfnamefont {M.}~\bibnamefont
  {Newman}}, \bibinfo {author} {\bibfnamefont {G.~N.}\ \bibnamefont {Kiladis}},
  \bibinfo {author} {\bibfnamefont {K.~M.}\ \bibnamefont {Weickmann}}, \bibinfo
  {author} {\bibfnamefont {F.~M.}\ \bibnamefont {Ralph}}, \ and\ \bibinfo
  {author} {\bibfnamefont {P.~D.}\ \bibnamefont {Sardeshmukh}},\ }\bibfield
  {title} {\enquote {\bibinfo {title} {Relative contributions of synoptic and
  low-frequency eddies to time-mean atmospheric moisture transport, including
  the role of atmospheric rivers},}\ }\href@noop {} {\bibfield  {journal}
  {\bibinfo  {journal} {Journal of climate}\ }\textbf {\bibinfo {volume}
  {25}},\ \bibinfo {pages} {7341--7361} (\bibinfo {year} {2012})}\BibitemShut
  {NoStop}%
\bibitem [{\citenamefont {Lavers}\ \emph {et~al.}(2015)\citenamefont {Lavers},
  \citenamefont {Ralph}, \citenamefont {Waliser}, \citenamefont {Gershunov},\
  and\ \citenamefont {Dettinger}}]{lavers2015climate}%
  \BibitemOpen
  \bibfield  {author} {\bibinfo {author} {\bibfnamefont {D.~A.}\ \bibnamefont
  {Lavers}}, \bibinfo {author} {\bibfnamefont {F.~M.}\ \bibnamefont {Ralph}},
  \bibinfo {author} {\bibfnamefont {D.~E.}\ \bibnamefont {Waliser}}, \bibinfo
  {author} {\bibfnamefont {A.}~\bibnamefont {Gershunov}}, \ and\ \bibinfo
  {author} {\bibfnamefont {M.~D.}\ \bibnamefont {Dettinger}},\ }\bibfield
  {title} {\enquote {\bibinfo {title} {Climate change intensification of
  horizontal water vapor transport in cmip5},}\ }\href@noop {} {\bibfield
  {journal} {\bibinfo  {journal} {Geophysical Research Letters}\ }\textbf
  {\bibinfo {volume} {42}},\ \bibinfo {pages} {5617--5625} (\bibinfo {year}
  {2015})}\BibitemShut {NoStop}%
\bibitem [{\citenamefont {Majda}(2012)}]{majda2012challenges}%
  \BibitemOpen
  \bibfield  {author} {\bibinfo {author} {\bibfnamefont {A.~J.}\ \bibnamefont
  {Majda}},\ }\bibfield  {title} {\enquote {\bibinfo {title} {Challenges in
  climate science and contemporary applied mathematics},}\ }\href@noop {}
  {\bibfield  {journal} {\bibinfo  {journal} {Communications on Pure and
  Applied Mathematics}\ }\textbf {\bibinfo {volume} {65}},\ \bibinfo {pages}
  {920--948} (\bibinfo {year} {2012})}\BibitemShut {NoStop}%
\bibitem [{\citenamefont {Majda}\ and\ \citenamefont
  {Chen}(2018)}]{majda2018model}%
  \BibitemOpen
  \bibfield  {author} {\bibinfo {author} {\bibfnamefont {A.~J.}\ \bibnamefont
  {Majda}}\ and\ \bibinfo {author} {\bibfnamefont {N.}~\bibnamefont {Chen}},\
  }\bibfield  {title} {\enquote {\bibinfo {title} {Model error, information
  barriers, state estimation and prediction in complex multiscale systems},}\
  }\href@noop {} {\bibfield  {journal} {\bibinfo  {journal} {Entropy}\ }\textbf
  {\bibinfo {volume} {20}},\ \bibinfo {pages} {644} (\bibinfo {year}
  {2018})}\BibitemShut {NoStop}%
\bibitem [{\citenamefont {Palmer}(2001)}]{palmer2001nonlinear}%
  \BibitemOpen
  \bibfield  {author} {\bibinfo {author} {\bibfnamefont {T.~N.}\ \bibnamefont
  {Palmer}},\ }\bibfield  {title} {\enquote {\bibinfo {title} {A nonlinear
  dynamical perspective on model error: A proposal for non-local
  stochastic-dynamic parametrization in weather and climate prediction
  models},}\ }\href@noop {} {\bibfield  {journal} {\bibinfo  {journal}
  {Quarterly Journal of the Royal Meteorological Society}\ }\textbf {\bibinfo
  {volume} {127}},\ \bibinfo {pages} {279--304} (\bibinfo {year}
  {2001})}\BibitemShut {NoStop}%
\bibitem [{\citenamefont {Majda}, \citenamefont {Franzke},\ and\ \citenamefont
  {Khouider}(2008)}]{majda2008applied}%
  \BibitemOpen
  \bibfield  {author} {\bibinfo {author} {\bibfnamefont {A.~J.}\ \bibnamefont
  {Majda}}, \bibinfo {author} {\bibfnamefont {C.}~\bibnamefont {Franzke}}, \
  and\ \bibinfo {author} {\bibfnamefont {B.}~\bibnamefont {Khouider}},\
  }\bibfield  {title} {\enquote {\bibinfo {title} {An applied mathematics
  perspective on stochastic modelling for climate},}\ }\href@noop {} {\bibfield
   {journal} {\bibinfo  {journal} {Philosophical Transactions of the Royal
  Society A: Mathematical, Physical and Engineering Sciences}\ }\textbf
  {\bibinfo {volume} {366}},\ \bibinfo {pages} {2427--2453} (\bibinfo {year}
  {2008})}\BibitemShut {NoStop}%
\bibitem [{\citenamefont {Edwards}(2011)}]{edwards2011history}%
  \BibitemOpen
  \bibfield  {author} {\bibinfo {author} {\bibfnamefont {P.~N.}\ \bibnamefont
  {Edwards}},\ }\bibfield  {title} {\enquote {\bibinfo {title} {History of
  climate modeling},}\ }\href@noop {} {\bibfield  {journal} {\bibinfo
  {journal} {Wiley Interdisciplinary Reviews: Climate Change}\ }\textbf
  {\bibinfo {volume} {2}},\ \bibinfo {pages} {128--139} (\bibinfo {year}
  {2011})}\BibitemShut {NoStop}%
\bibitem [{\citenamefont {Franzke}, \citenamefont {Majda},\ and\ \citenamefont
  {Vanden-Eijnden}(2005)}]{franzke2005low}%
  \BibitemOpen
  \bibfield  {author} {\bibinfo {author} {\bibfnamefont {C.}~\bibnamefont
  {Franzke}}, \bibinfo {author} {\bibfnamefont {A.~J.}\ \bibnamefont {Majda}},
  \ and\ \bibinfo {author} {\bibfnamefont {E.}~\bibnamefont {Vanden-Eijnden}},\
  }\bibfield  {title} {\enquote {\bibinfo {title} {Low-order stochastic mode
  reduction for a realistic barotropic model climate},}\ }\href@noop {}
  {\bibfield  {journal} {\bibinfo  {journal} {Journal of the atmospheric
  sciences}\ }\textbf {\bibinfo {volume} {62}},\ \bibinfo {pages} {1722--1745}
  (\bibinfo {year} {2005})}\BibitemShut {NoStop}%
\bibitem [{\citenamefont {Mou}, \citenamefont {Chen},\ and\ \citenamefont
  {Iliescu}(2023)}]{mou2023efficient}%
  \BibitemOpen
  \bibfield  {author} {\bibinfo {author} {\bibfnamefont {C.}~\bibnamefont
  {Mou}}, \bibinfo {author} {\bibfnamefont {N.}~\bibnamefont {Chen}}, \ and\
  \bibinfo {author} {\bibfnamefont {T.}~\bibnamefont {Iliescu}},\ }\bibfield
  {title} {\enquote {\bibinfo {title} {An efficient data-driven multiscale
  stochastic reduced order modeling framework for complex systems},}\
  }\href@noop {} {\bibfield  {journal} {\bibinfo  {journal} {Journal of
  Computational Physics}\ }\textbf {\bibinfo {volume} {493}},\ \bibinfo {pages}
  {112450} (\bibinfo {year} {2023})}\BibitemShut {NoStop}%
\bibitem [{\citenamefont {Majda}, \citenamefont {Timofeyev},\ and\
  \citenamefont {Vanden~Eijnden}(2001)}]{majda2001mathematical}%
  \BibitemOpen
  \bibfield  {author} {\bibinfo {author} {\bibfnamefont {A.~J.}\ \bibnamefont
  {Majda}}, \bibinfo {author} {\bibfnamefont {I.}~\bibnamefont {Timofeyev}}, \
  and\ \bibinfo {author} {\bibfnamefont {E.}~\bibnamefont {Vanden~Eijnden}},\
  }\bibfield  {title} {\enquote {\bibinfo {title} {A mathematical framework for
  stochastic climate models},}\ }\href@noop {} {\bibfield  {journal} {\bibinfo
  {journal} {Communications on Pure and Applied Mathematics: A Journal Issued
  by the Courant Institute of Mathematical Sciences}\ }\textbf {\bibinfo
  {volume} {54}},\ \bibinfo {pages} {891--974} (\bibinfo {year}
  {2001})}\BibitemShut {NoStop}%
\bibitem [{\citenamefont {Franzke}\ \emph {et~al.}(2015)\citenamefont
  {Franzke}, \citenamefont {O'Kane}, \citenamefont {Berner}, \citenamefont
  {Williams},\ and\ \citenamefont {Lucarini}}]{franzke2015stochastic}%
  \BibitemOpen
  \bibfield  {author} {\bibinfo {author} {\bibfnamefont {C.~L.}\ \bibnamefont
  {Franzke}}, \bibinfo {author} {\bibfnamefont {T.~J.}\ \bibnamefont {O'Kane}},
  \bibinfo {author} {\bibfnamefont {J.}~\bibnamefont {Berner}}, \bibinfo
  {author} {\bibfnamefont {P.~D.}\ \bibnamefont {Williams}}, \ and\ \bibinfo
  {author} {\bibfnamefont {V.}~\bibnamefont {Lucarini}},\ }\bibfield  {title}
  {\enquote {\bibinfo {title} {Stochastic climate theory and modeling},}\
  }\href@noop {} {\bibfield  {journal} {\bibinfo  {journal} {Wiley
  Interdisciplinary Reviews: Climate Change}\ }\textbf {\bibinfo {volume}
  {6}},\ \bibinfo {pages} {63--78} (\bibinfo {year} {2015})}\BibitemShut
  {NoStop}%
\bibitem [{\citenamefont {Imkeller}\ and\ \citenamefont
  {Von~Storch}(2001)}]{imkeller2001stochastic}%
  \BibitemOpen
  \bibfield  {author} {\bibinfo {author} {\bibfnamefont {P.}~\bibnamefont
  {Imkeller}}\ and\ \bibinfo {author} {\bibfnamefont {J.-S.}\ \bibnamefont
  {Von~Storch}},\ }\href@noop {} {\emph {\bibinfo {title} {Stochastic climate
  models}}},\ Vol.~\bibinfo {volume} {49}\ (\bibinfo  {publisher} {Springer
  Science \& Business Media},\ \bibinfo {year} {2001})\BibitemShut {NoStop}%
\bibitem [{\citenamefont {Majda}, \citenamefont {Timofeyev},\ and\
  \citenamefont {Vanden~Eijnden}(1999)}]{majda1999models}%
  \BibitemOpen
  \bibfield  {author} {\bibinfo {author} {\bibfnamefont {A.~J.}\ \bibnamefont
  {Majda}}, \bibinfo {author} {\bibfnamefont {I.}~\bibnamefont {Timofeyev}}, \
  and\ \bibinfo {author} {\bibfnamefont {E.}~\bibnamefont {Vanden~Eijnden}},\
  }\bibfield  {title} {\enquote {\bibinfo {title} {Models for stochastic
  climate prediction},}\ }\href@noop {} {\bibfield  {journal} {\bibinfo
  {journal} {Proceedings of the National Academy of Sciences}\ }\textbf
  {\bibinfo {volume} {96}},\ \bibinfo {pages} {14687--14691} (\bibinfo {year}
  {1999})}\BibitemShut {NoStop}%
\bibitem [{\citenamefont {Majda}, \citenamefont {Timofeyev},\ and\
  \citenamefont {Vanden-Eijnden}(2003)}]{majda2003systematic}%
  \BibitemOpen
  \bibfield  {author} {\bibinfo {author} {\bibfnamefont {A.~J.}\ \bibnamefont
  {Majda}}, \bibinfo {author} {\bibfnamefont {I.}~\bibnamefont {Timofeyev}}, \
  and\ \bibinfo {author} {\bibfnamefont {E.}~\bibnamefont {Vanden-Eijnden}},\
  }\bibfield  {title} {\enquote {\bibinfo {title} {Systematic strategies for
  stochastic mode reduction in climate},}\ }\href@noop {} {\bibfield  {journal}
  {\bibinfo  {journal} {Journal of the Atmospheric Sciences}\ }\textbf
  {\bibinfo {volume} {60}},\ \bibinfo {pages} {1705--1722} (\bibinfo {year}
  {2003})}\BibitemShut {NoStop}%
\bibitem [{\citenamefont {Chen}\ and\ \citenamefont
  {Majda}(2018)}]{chen2018conditional}%
  \BibitemOpen
  \bibfield  {author} {\bibinfo {author} {\bibfnamefont {N.}~\bibnamefont
  {Chen}}\ and\ \bibinfo {author} {\bibfnamefont {A.~J.}\ \bibnamefont
  {Majda}},\ }\bibfield  {title} {\enquote {\bibinfo {title} {Conditional
  {G}aussian systems for multiscale nonlinear stochastic systems: Prediction,
  state estimation and uncertainty quantification},}\ }\href@noop {} {\bibfield
   {journal} {\bibinfo  {journal} {Entropy}\ }\textbf {\bibinfo {volume}
  {20}},\ \bibinfo {pages} {509} (\bibinfo {year} {2018})}\BibitemShut
  {NoStop}%
\bibitem [{\citenamefont {Palmer}(2019)}]{palmer2019stochastic}%
  \BibitemOpen
  \bibfield  {author} {\bibinfo {author} {\bibfnamefont {T.}~\bibnamefont
  {Palmer}},\ }\bibfield  {title} {\enquote {\bibinfo {title} {Stochastic
  weather and climate models},}\ }\href@noop {} {\bibfield  {journal} {\bibinfo
   {journal} {Nature Reviews Physics}\ }\textbf {\bibinfo {volume} {1}},\
  \bibinfo {pages} {463--471} (\bibinfo {year} {2019})}\BibitemShut {NoStop}%
\bibitem [{\citenamefont {Horenko}\ \emph {et~al.}(2008)\citenamefont
  {Horenko}, \citenamefont {Klein}, \citenamefont {Dolaptchiev},\ and\
  \citenamefont {Sch{\"u}tte}}]{horenko2008automated}%
  \BibitemOpen
  \bibfield  {author} {\bibinfo {author} {\bibfnamefont {I.}~\bibnamefont
  {Horenko}}, \bibinfo {author} {\bibfnamefont {R.}~\bibnamefont {Klein}},
  \bibinfo {author} {\bibfnamefont {S.}~\bibnamefont {Dolaptchiev}}, \ and\
  \bibinfo {author} {\bibfnamefont {C.}~\bibnamefont {Sch{\"u}tte}},\
  }\bibfield  {title} {\enquote {\bibinfo {title} {Automated generation of
  reduced stochastic weather models i: simultaneous dimension and model
  reduction for time series analysis},}\ }\href@noop {} {\bibfield  {journal}
  {\bibinfo  {journal} {Multiscale Modeling \& Simulation}\ }\textbf {\bibinfo
  {volume} {6}},\ \bibinfo {pages} {1125--1145} (\bibinfo {year}
  {2008})}\BibitemShut {NoStop}%
\bibitem [{\citenamefont {Majda}, \citenamefont {Franzke},\ and\ \citenamefont
  {Crommelin}(2009)}]{majda2009normal}%
  \BibitemOpen
  \bibfield  {author} {\bibinfo {author} {\bibfnamefont {A.~J.}\ \bibnamefont
  {Majda}}, \bibinfo {author} {\bibfnamefont {C.}~\bibnamefont {Franzke}}, \
  and\ \bibinfo {author} {\bibfnamefont {D.}~\bibnamefont {Crommelin}},\
  }\bibfield  {title} {\enquote {\bibinfo {title} {Normal forms for reduced
  stochastic climate models},}\ }\href@noop {} {\bibfield  {journal} {\bibinfo
  {journal} {Proceedings of the National Academy of Sciences}\ }\textbf
  {\bibinfo {volume} {106}},\ \bibinfo {pages} {3649--3653} (\bibinfo {year}
  {2009})}\BibitemShut {NoStop}%
\bibitem [{\citenamefont {Smith}\ and\ \citenamefont
  {Stechmann}(2017)}]{smith2017precipitating}%
  \BibitemOpen
  \bibfield  {author} {\bibinfo {author} {\bibfnamefont {L.~M.}\ \bibnamefont
  {Smith}}\ and\ \bibinfo {author} {\bibfnamefont {S.~N.}\ \bibnamefont
  {Stechmann}},\ }\bibfield  {title} {\enquote {\bibinfo {title} {Precipitating
  quasigeostrophic equations and potential vorticity inversion with phase
  changes},}\ }\href@noop {} {\bibfield  {journal} {\bibinfo  {journal}
  {Journal of the Atmospheric Sciences}\ }\textbf {\bibinfo {volume} {74}},\
  \bibinfo {pages} {3285--3303} (\bibinfo {year} {2017})}\BibitemShut {NoStop}%
\bibitem [{\citenamefont {Law}, \citenamefont {Stuart},\ and\ \citenamefont
  {Zygalakis}(2015)}]{law2015data}%
  \BibitemOpen
  \bibfield  {author} {\bibinfo {author} {\bibfnamefont {K.}~\bibnamefont
  {Law}}, \bibinfo {author} {\bibfnamefont {A.}~\bibnamefont {Stuart}}, \ and\
  \bibinfo {author} {\bibfnamefont {K.}~\bibnamefont {Zygalakis}},\ }\bibfield
  {title} {\enquote {\bibinfo {title} {Data assimilation},}\ }\href@noop {}
  {\bibfield  {journal} {\bibinfo  {journal} {Cham, Switzerland: Springer}\
  }\textbf {\bibinfo {volume} {214}},\ \bibinfo {pages} {52} (\bibinfo {year}
  {2015})}\BibitemShut {NoStop}%
\bibitem [{\citenamefont {Asch}, \citenamefont {Bocquet},\ and\ \citenamefont
  {Nodet}(2016)}]{asch2016data}%
  \BibitemOpen
  \bibfield  {author} {\bibinfo {author} {\bibfnamefont {M.}~\bibnamefont
  {Asch}}, \bibinfo {author} {\bibfnamefont {M.}~\bibnamefont {Bocquet}}, \
  and\ \bibinfo {author} {\bibfnamefont {M.}~\bibnamefont {Nodet}},\
  }\href@noop {} {\emph {\bibinfo {title} {Data assimilation: methods,
  algorithms, and applications}}}\ (\bibinfo  {publisher} {SIAM},\ \bibinfo
  {year} {2016})\BibitemShut {NoStop}%
\bibitem [{\citenamefont {Charney}(1948)}]{charney1948scale}%
  \BibitemOpen
  \bibfield  {author} {\bibinfo {author} {\bibfnamefont {J.~G.}\ \bibnamefont
  {Charney}},\ }\bibfield  {title} {\enquote {\bibinfo {title} {On the scale of
  atmospheric motions},}\ }in\ \href@noop {} {\emph {\bibinfo {booktitle} {The
  Atmosphere—A Challenge: The Science of Jule Gregory Charney}}}\ (\bibinfo
  {publisher} {Springer},\ \bibinfo {year} {1948})\ pp.\ \bibinfo {pages}
  {251--265}\BibitemShut {NoStop}%
\bibitem [{\citenamefont {Majda}(2003)}]{majda2003introduction}%
  \BibitemOpen
  \bibfield  {author} {\bibinfo {author} {\bibfnamefont {A.}~\bibnamefont
  {Majda}},\ }\href@noop {} {\emph {\bibinfo {title} {Introduction to PDEs and
  Waves for the Atmosphere and Ocean}}},\ Vol.~\bibinfo {volume} {9}\ (\bibinfo
   {publisher} {American Mathematical Soc.},\ \bibinfo {year}
  {2003})\BibitemShut {NoStop}%
\bibitem [{\citenamefont {Vallis}(2017)}]{vallis2017atmospheric}%
  \BibitemOpen
  \bibfield  {author} {\bibinfo {author} {\bibfnamefont {G.~K.}\ \bibnamefont
  {Vallis}},\ }\href@noop {} {\emph {\bibinfo {title} {Atmospheric and oceanic
  fluid dynamics}}}\ (\bibinfo  {publisher} {Cambridge University Press},\
  \bibinfo {year} {2017})\BibitemShut {NoStop}%
\bibitem [{\citenamefont {Emanuel}, \citenamefont {Fantini},\ and\
  \citenamefont {Thorpe}(1987)}]{emanuel1987baroclinic}%
  \BibitemOpen
  \bibfield  {author} {\bibinfo {author} {\bibfnamefont {K.~A.}\ \bibnamefont
  {Emanuel}}, \bibinfo {author} {\bibfnamefont {M.}~\bibnamefont {Fantini}}, \
  and\ \bibinfo {author} {\bibfnamefont {A.~J.}\ \bibnamefont {Thorpe}},\
  }\bibfield  {title} {\enquote {\bibinfo {title} {Baroclinic instability in an
  environment of small stability to slantwise moist convection},}\ }\href@noop
  {} {\bibfield  {journal} {\bibinfo  {journal} {J. Atmos. Sci}\ }\textbf
  {\bibinfo {volume} {44}},\ \bibinfo {pages} {1559--1573} (\bibinfo {year}
  {1987})}\BibitemShut {NoStop}%
\bibitem [{\citenamefont {Lapeyre}\ and\ \citenamefont
  {Held}(2004)}]{lapeyre2004role}%
  \BibitemOpen
  \bibfield  {author} {\bibinfo {author} {\bibfnamefont {G.}~\bibnamefont
  {Lapeyre}}\ and\ \bibinfo {author} {\bibfnamefont {I.}~\bibnamefont {Held}},\
  }\bibfield  {title} {\enquote {\bibinfo {title} {The role of moisture in the
  dynamics and energetics of turbulent baroclinic eddies},}\ }\href@noop {}
  {\bibfield  {journal} {\bibinfo  {journal} {Journal of the atmospheric
  sciences}\ }\textbf {\bibinfo {volume} {61}},\ \bibinfo {pages} {1693--1710}
  (\bibinfo {year} {2004})}\BibitemShut {NoStop}%
\bibitem [{\citenamefont {Ertel}(1942)}]{ertel1942neuer}%
  \BibitemOpen
  \bibfield  {author} {\bibinfo {author} {\bibfnamefont {H.}~\bibnamefont
  {Ertel}},\ }\bibfield  {title} {\enquote {\bibinfo {title} {Ein neuer
  hydrofynamischer wirbelsatz.}}\ }\href@noop {} {\bibfield  {journal}
  {\bibinfo  {journal} {Meteor. Z.}\ }\textbf {\bibinfo {volume} {59}},\
  \bibinfo {pages} {277--282} (\bibinfo {year} {1942})}\BibitemShut {NoStop}%
\bibitem [{\citenamefont {Hoskins}, \citenamefont {McIntyre},\ and\
  \citenamefont {Robertson}(1985)}]{hoskins1985use}%
  \BibitemOpen
  \bibfield  {author} {\bibinfo {author} {\bibfnamefont {B.~J.}\ \bibnamefont
  {Hoskins}}, \bibinfo {author} {\bibfnamefont {M.~E.}\ \bibnamefont
  {McIntyre}}, \ and\ \bibinfo {author} {\bibfnamefont {A.~W.}\ \bibnamefont
  {Robertson}},\ }\bibfield  {title} {\enquote {\bibinfo {title} {On the use
  and significance of isentropic potential vorticity maps},}\ }\href@noop {}
  {\bibfield  {journal} {\bibinfo  {journal} {Quarterly Journal of the Royal
  Meteorological Society}\ }\textbf {\bibinfo {volume} {111}},\ \bibinfo
  {pages} {877--946} (\bibinfo {year} {1985})}\BibitemShut {NoStop}%
\bibitem [{\citenamefont {Edwards}, \citenamefont {Smith},\ and\ \citenamefont
  {Stechmann}(2020{\natexlab{a}})}]{edwards2020spectra}%
  \BibitemOpen
  \bibfield  {author} {\bibinfo {author} {\bibfnamefont {T.~K.}\ \bibnamefont
  {Edwards}}, \bibinfo {author} {\bibfnamefont {L.~M.}\ \bibnamefont {Smith}},
  \ and\ \bibinfo {author} {\bibfnamefont {S.~N.}\ \bibnamefont {Stechmann}},\
  }\bibfield  {title} {\enquote {\bibinfo {title} {Spectra of atmospheric water
  in precipitating quasi-geostrophic turbulence},}\ }\href@noop {} {\bibfield
  {journal} {\bibinfo  {journal} {Geophysical \& Astrophysical Fluid Dynamics}\
  }\textbf {\bibinfo {volume} {114}},\ \bibinfo {pages} {715--741} (\bibinfo
  {year} {2020}{\natexlab{a}})}\BibitemShut {NoStop}%
\bibitem [{\citenamefont {Hu}\ \emph {et~al.}(2021)\citenamefont {Hu},
  \citenamefont {Edwards}, \citenamefont {Smith},\ and\ \citenamefont
  {Stechmann}}]{hu2021initial}%
  \BibitemOpen
  \bibfield  {author} {\bibinfo {author} {\bibfnamefont {R.}~\bibnamefont
  {Hu}}, \bibinfo {author} {\bibfnamefont {T.~K.}\ \bibnamefont {Edwards}},
  \bibinfo {author} {\bibfnamefont {L.~M.}\ \bibnamefont {Smith}}, \ and\
  \bibinfo {author} {\bibfnamefont {S.~N.}\ \bibnamefont {Stechmann}},\
  }\bibfield  {title} {\enquote {\bibinfo {title} {Initial investigations of
  precipitating quasi-geostrophic turbulence with phase changes},}\ }\href@noop
  {} {\bibfield  {journal} {\bibinfo  {journal} {Research in the Mathematical
  Sciences}\ }\textbf {\bibinfo {volume} {8}},\ \bibinfo {pages} {1--25}
  (\bibinfo {year} {2021})}\BibitemShut {NoStop}%
\bibitem [{\citenamefont {Mou}, \citenamefont {Smith},\ and\ \citenamefont
  {Chen}(2023)}]{mou2023combining}%
  \BibitemOpen
  \bibfield  {author} {\bibinfo {author} {\bibfnamefont {C.}~\bibnamefont
  {Mou}}, \bibinfo {author} {\bibfnamefont {L.~M.}\ \bibnamefont {Smith}}, \
  and\ \bibinfo {author} {\bibfnamefont {N.}~\bibnamefont {Chen}},\ }\bibfield
  {title} {\enquote {\bibinfo {title} {Combining stochastic parameterized
  reduced-order models with machine learning for data assimilation and
  uncertainty quantification with partial observations},}\ }\href@noop {}
  {\bibfield  {journal} {\bibinfo  {journal} {Journal of Advances in Modeling
  Earth Systems}\ }\textbf {\bibinfo {volume} {15}},\ \bibinfo {pages}
  {e2022MS003597} (\bibinfo {year} {2023})}\BibitemShut {NoStop}%
\bibitem [{\citenamefont {Gardiner}(2004)}]{gardiner2004handbook}%
  \BibitemOpen
  \bibfield  {author} {\bibinfo {author} {\bibfnamefont {C.~W.}\ \bibnamefont
  {Gardiner}},\ }\href@noop {} {\enquote {\bibinfo {title} {Handbook of
  stochastic methods for physics, chemistry and the natural sciences, vol. 13
  of springer series in synergetics},}\ } (\bibinfo {year} {2004})\BibitemShut
  {NoStop}%
\bibitem [{\citenamefont {Katsoulakis}, \citenamefont {Majda},\ and\
  \citenamefont {Vlachos}(2003)}]{katsoulakis2003coarse}%
  \BibitemOpen
  \bibfield  {author} {\bibinfo {author} {\bibfnamefont {M.~A.}\ \bibnamefont
  {Katsoulakis}}, \bibinfo {author} {\bibfnamefont {A.~J.}\ \bibnamefont
  {Majda}}, \ and\ \bibinfo {author} {\bibfnamefont {D.~G.}\ \bibnamefont
  {Vlachos}},\ }\bibfield  {title} {\enquote {\bibinfo {title} {Coarse-grained
  stochastic processes for microscopic lattice systems},}\ }\href@noop {}
  {\bibfield  {journal} {\bibinfo  {journal} {Proceedings of the National
  Academy of Sciences}\ }\textbf {\bibinfo {volume} {100}},\ \bibinfo {pages}
  {782--787} (\bibinfo {year} {2003})}\BibitemShut {NoStop}%
\bibitem [{\citenamefont {Khouider}, \citenamefont {Biello},\ and\
  \citenamefont {Majda}(2010)}]{khouider2010stochastic}%
  \BibitemOpen
  \bibfield  {author} {\bibinfo {author} {\bibfnamefont {B.}~\bibnamefont
  {Khouider}}, \bibinfo {author} {\bibfnamefont {J.}~\bibnamefont {Biello}}, \
  and\ \bibinfo {author} {\bibfnamefont {A.~J.}\ \bibnamefont {Majda}},\
  }\bibfield  {title} {\enquote {\bibinfo {title} {A stochastic multicloud
  model for tropical convection},}\ }\href@noop {} {\bibfield  {journal}
  {\bibinfo  {journal} {Commun. Math. Sci.}\ }\textbf {\bibinfo {volume} {8}},\
  \bibinfo {pages} {187--216} (\bibinfo {year} {2010})}\BibitemShut {NoStop}%
\bibitem [{\citenamefont {Salmon}(1980)}]{salmon1980baroclinic}%
  \BibitemOpen
  \bibfield  {author} {\bibinfo {author} {\bibfnamefont {R.}~\bibnamefont
  {Salmon}},\ }\bibfield  {title} {\enquote {\bibinfo {title} {Baroclinic
  instability and geostrophic turbulence},}\ }\href@noop {} {\bibfield
  {journal} {\bibinfo  {journal} {Geophysical \& Astrophysical Fluid Dynamics}\
  }\textbf {\bibinfo {volume} {15}},\ \bibinfo {pages} {167--211} (\bibinfo
  {year} {1980})}\BibitemShut {NoStop}%
\bibitem [{\citenamefont {Qi}\ and\ \citenamefont {Majda}(2016)}]{qi2016low}%
  \BibitemOpen
  \bibfield  {author} {\bibinfo {author} {\bibfnamefont {D.}~\bibnamefont
  {Qi}}\ and\ \bibinfo {author} {\bibfnamefont {A.~J.}\ \bibnamefont {Majda}},\
  }\bibfield  {title} {\enquote {\bibinfo {title} {Low-dimensional
  reduced-order models for statistical response and uncertainty quantification:
  Two-layer baroclinic turbulence},}\ }\href@noop {} {\bibfield  {journal}
  {\bibinfo  {journal} {Journal of the Atmospheric Sciences}\ }\textbf
  {\bibinfo {volume} {73}},\ \bibinfo {pages} {4609--4639} (\bibinfo {year}
  {2016})}\BibitemShut {NoStop}%
\bibitem [{\citenamefont {Hernandez-Duenas}\ \emph {et~al.}(2013)\citenamefont
  {Hernandez-Duenas}, \citenamefont {Majda}, \citenamefont {Smith},\ and\
  \citenamefont {Stechmann}}]{hernandez2013minimal}%
  \BibitemOpen
  \bibfield  {author} {\bibinfo {author} {\bibfnamefont {G.}~\bibnamefont
  {Hernandez-Duenas}}, \bibinfo {author} {\bibfnamefont {A.~J.}\ \bibnamefont
  {Majda}}, \bibinfo {author} {\bibfnamefont {L.~M.}\ \bibnamefont {Smith}}, \
  and\ \bibinfo {author} {\bibfnamefont {S.~N.}\ \bibnamefont {Stechmann}},\
  }\bibfield  {title} {\enquote {\bibinfo {title} {Minimal models for
  precipitating turbulent convection},}\ }\href@noop {} {\bibfield  {journal}
  {\bibinfo  {journal} {Journal of Fluid Mechanics}\ }\textbf {\bibinfo
  {volume} {717}},\ \bibinfo {pages} {576--611} (\bibinfo {year}
  {2013})}\BibitemShut {NoStop}%
\bibitem [{\citenamefont {Zhang}, \citenamefont {Smith},\ and\ \citenamefont
  {Stechmann}(2021{\natexlab{a}})}]{zhang2021effects}%
  \BibitemOpen
  \bibfield  {author} {\bibinfo {author} {\bibfnamefont {Y.}~\bibnamefont
  {Zhang}}, \bibinfo {author} {\bibfnamefont {L.~M.}\ \bibnamefont {Smith}}, \
  and\ \bibinfo {author} {\bibfnamefont {S.~N.}\ \bibnamefont {Stechmann}},\
  }\bibfield  {title} {\enquote {\bibinfo {title} {Effects of clouds and phase
  changes on fast-wave averaging: a numerical assessment},}\ }\href@noop {}
  {\bibfield  {journal} {\bibinfo  {journal} {Journal of Fluid Mechanics}\
  }\textbf {\bibinfo {volume} {920}},\ \bibinfo {pages} {A49} (\bibinfo {year}
  {2021}{\natexlab{a}})}\BibitemShut {NoStop}%
\bibitem [{\citenamefont {Remond-Tiedrez}, \citenamefont {Smith},\ and\
  \citenamefont {Stechmann}(2024)}]{remond2024nonlinear}%
  \BibitemOpen
  \bibfield  {author} {\bibinfo {author} {\bibfnamefont {A.}~\bibnamefont
  {Remond-Tiedrez}}, \bibinfo {author} {\bibfnamefont {L.~M.}\ \bibnamefont
  {Smith}}, \ and\ \bibinfo {author} {\bibfnamefont {S.~N.}\ \bibnamefont
  {Stechmann}},\ }\bibfield  {title} {\enquote {\bibinfo {title} {A nonlinear
  elliptic pde from atmospheric science: well-posedness and regularity at cloud
  edge},}\ }\href@noop {} {\bibfield  {journal} {\bibinfo  {journal} {Journal
  of Mathematical Fluid Mechanics}\ }\textbf {\bibinfo {volume} {26}},\
  \bibinfo {pages} {30} (\bibinfo {year} {2024})}\BibitemShut {NoStop}%
\bibitem [{\citenamefont {Wetzel}, \citenamefont {Smith},\ and\ \citenamefont
  {Stechmann}(2017)}]{wetzel2017moisture}%
  \BibitemOpen
  \bibfield  {author} {\bibinfo {author} {\bibfnamefont {A.~N.}\ \bibnamefont
  {Wetzel}}, \bibinfo {author} {\bibfnamefont {L.~M.}\ \bibnamefont {Smith}}, \
  and\ \bibinfo {author} {\bibfnamefont {S.~N.}\ \bibnamefont {Stechmann}},\
  }\bibfield  {title} {\enquote {\bibinfo {title} {Moisture transport due to
  baroclinic waves: Linear analysis of precipitating quasi-geostrophic
  dynamics},}\ }\href@noop {} {\bibfield  {journal} {\bibinfo  {journal}
  {Mathematics of Climate and Weather Forecasting}\ }\textbf {\bibinfo {volume}
  {3}},\ \bibinfo {pages} {28--50} (\bibinfo {year} {2017})}\BibitemShut
  {NoStop}%
\bibitem [{\citenamefont {Wetzel}, \citenamefont {Smith},\ and\ \citenamefont
  {Stechmann}(2019)}]{wetzel2019discontinuous}%
  \BibitemOpen
  \bibfield  {author} {\bibinfo {author} {\bibfnamefont {A.~N.}\ \bibnamefont
  {Wetzel}}, \bibinfo {author} {\bibfnamefont {L.~M.}\ \bibnamefont {Smith}}, \
  and\ \bibinfo {author} {\bibfnamefont {S.~N.}\ \bibnamefont {Stechmann}},\
  }\bibfield  {title} {\enquote {\bibinfo {title} {Discontinuous fronts as
  exact solutions to precipitating quasi-geostrophic equations},}\ }\href@noop
  {} {\bibfield  {journal} {\bibinfo  {journal} {SIAM Journal on Applied
  Mathematics}\ }\textbf {\bibinfo {volume} {79}},\ \bibinfo {pages}
  {1341--1366} (\bibinfo {year} {2019})}\BibitemShut {NoStop}%
\bibitem [{\citenamefont {Edwards}, \citenamefont {Smith},\ and\ \citenamefont
  {Stechmann}(2020{\natexlab{b}})}]{edwards2020atmospheric}%
  \BibitemOpen
  \bibfield  {author} {\bibinfo {author} {\bibfnamefont {T.~K.}\ \bibnamefont
  {Edwards}}, \bibinfo {author} {\bibfnamefont {L.~M.}\ \bibnamefont {Smith}},
  \ and\ \bibinfo {author} {\bibfnamefont {S.~N.}\ \bibnamefont {Stechmann}},\
  }\bibfield  {title} {\enquote {\bibinfo {title} {Atmospheric rivers and water
  fluxes in precipitating quasi-geostrophic turbulence},}\ }\href@noop {}
  {\bibfield  {journal} {\bibinfo  {journal} {Quarterly Journal of the Royal
  Meteorological Society}\ }\textbf {\bibinfo {volume} {146}},\ \bibinfo
  {pages} {1960--1975} (\bibinfo {year} {2020}{\natexlab{b}})}\BibitemShut
  {NoStop}%
\bibitem [{\citenamefont {Treguier}\ and\ \citenamefont
  {Hua}(1987)}]{treguier1987oceanic}%
  \BibitemOpen
  \bibfield  {author} {\bibinfo {author} {\bibfnamefont {A.~M.}\ \bibnamefont
  {Treguier}}\ and\ \bibinfo {author} {\bibfnamefont {B.~L.}\ \bibnamefont
  {Hua}},\ }\bibfield  {title} {\enquote {\bibinfo {title} {Oceanic
  quasi-geostrophic turbulence forced by stochastic wind fluctuations},}\
  }\href@noop {} {\bibfield  {journal} {\bibinfo  {journal} {Journal of
  physical oceanography}\ }\textbf {\bibinfo {volume} {17}},\ \bibinfo {pages}
  {397--411} (\bibinfo {year} {1987})}\BibitemShut {NoStop}%
\bibitem [{\citenamefont {Martius}\ \emph {et~al.}(2006)\citenamefont
  {Martius}, \citenamefont {Zenklusen}, \citenamefont {Schwierz},\ and\
  \citenamefont {Davies}}]{martius2006episodes}%
  \BibitemOpen
  \bibfield  {author} {\bibinfo {author} {\bibfnamefont {O.}~\bibnamefont
  {Martius}}, \bibinfo {author} {\bibfnamefont {E.}~\bibnamefont {Zenklusen}},
  \bibinfo {author} {\bibfnamefont {C.}~\bibnamefont {Schwierz}}, \ and\
  \bibinfo {author} {\bibfnamefont {H.~C.}\ \bibnamefont {Davies}},\ }\bibfield
   {title} {\enquote {\bibinfo {title} {Episodes of alpine heavy precipitation
  with an overlying elongated stratospheric intrusion: A climatology},}\
  }\href@noop {} {\bibfield  {journal} {\bibinfo  {journal} {International
  Journal of Climatology: A Journal of the Royal Meteorological Society}\
  }\textbf {\bibinfo {volume} {26}},\ \bibinfo {pages} {1149--1164} (\bibinfo
  {year} {2006})}\BibitemShut {NoStop}%
\bibitem [{\citenamefont {Roberts}(2000)}]{roberts2000relationship}%
  \BibitemOpen
  \bibfield  {author} {\bibinfo {author} {\bibfnamefont {N.~M.}\ \bibnamefont
  {Roberts}},\ }\href@noop {} {\emph {\bibinfo {title} {The relationship
  between water vapour imagery and thunderstorms}}}\ (\bibinfo  {publisher}
  {Joint Centre for Mesoscale Meteorology},\ \bibinfo {year}
  {2000})\BibitemShut {NoStop}%
\bibitem [{\citenamefont {Chagnon}\ and\ \citenamefont
  {Gray}(2009)}]{chagnon2009horizontal}%
  \BibitemOpen
  \bibfield  {author} {\bibinfo {author} {\bibfnamefont {J.~M.}\ \bibnamefont
  {Chagnon}}\ and\ \bibinfo {author} {\bibfnamefont {S.~L.}\ \bibnamefont
  {Gray}},\ }\bibfield  {title} {\enquote {\bibinfo {title} {Horizontal
  potential vorticity dipoles on the convective storm scale},}\ }\href@noop {}
  {\bibfield  {journal} {\bibinfo  {journal} {Quarterly Journal of the Royal
  Meteorological Society: A journal of the atmospheric sciences, applied
  meteorology and physical oceanography}\ }\textbf {\bibinfo {volume} {135}},\
  \bibinfo {pages} {1392--1408} (\bibinfo {year} {2009})}\BibitemShut {NoStop}%
\bibitem [{\citenamefont {Schubert}\ \emph {et~al.}(2004)\citenamefont
  {Schubert}, \citenamefont {Ruprecht}, \citenamefont {Hertenstein},
  \citenamefont {Ferreira}, \citenamefont {Taft}, \citenamefont {Rozoff},
  \citenamefont {Ciesielski},\ and\ \citenamefont {Kuo}}]{schubert2004english}%
  \BibitemOpen
  \bibfield  {author} {\bibinfo {author} {\bibfnamefont {W.}~\bibnamefont
  {Schubert}}, \bibinfo {author} {\bibfnamefont {E.}~\bibnamefont {Ruprecht}},
  \bibinfo {author} {\bibfnamefont {R.}~\bibnamefont {Hertenstein}}, \bibinfo
  {author} {\bibfnamefont {R.~N.}\ \bibnamefont {Ferreira}}, \bibinfo {author}
  {\bibfnamefont {R.}~\bibnamefont {Taft}}, \bibinfo {author} {\bibfnamefont
  {C.}~\bibnamefont {Rozoff}}, \bibinfo {author} {\bibfnamefont
  {P.}~\bibnamefont {Ciesielski}}, \ and\ \bibinfo {author} {\bibfnamefont
  {H.-C.}\ \bibnamefont {Kuo}},\ }\bibfield  {title} {\enquote {\bibinfo
  {title} {English translations of twenty-one of ertel's papers on geophysical
  fluid dynamics},}\ }\href@noop {} {\bibfield  {journal} {\bibinfo  {journal}
  {Meteorology Z}\ ,\ \bibinfo {pages} {527--576}} (\bibinfo {year}
  {2004})}\BibitemShut {NoStop}%
\bibitem [{\citenamefont {Babaud}\ \emph {et~al.}(1986)\citenamefont {Babaud},
  \citenamefont {Witkin}, \citenamefont {Baudin},\ and\ \citenamefont
  {Duda}}]{babaud1986uniqueness}%
  \BibitemOpen
  \bibfield  {author} {\bibinfo {author} {\bibfnamefont {J.}~\bibnamefont
  {Babaud}}, \bibinfo {author} {\bibfnamefont {A.~P.}\ \bibnamefont {Witkin}},
  \bibinfo {author} {\bibfnamefont {M.}~\bibnamefont {Baudin}}, \ and\ \bibinfo
  {author} {\bibfnamefont {R.~O.}\ \bibnamefont {Duda}},\ }\bibfield  {title}
  {\enquote {\bibinfo {title} {Uniqueness of the gaussian kernel for
  scale-space filtering},}\ }\href@noop {} {\bibfield  {journal} {\bibinfo
  {journal} {IEEE transactions on pattern analysis and machine intelligence}\
  ,\ \bibinfo {pages} {26--33}} (\bibinfo {year} {1986})}\BibitemShut {NoStop}%
\bibitem [{\citenamefont {Wand}\ and\ \citenamefont
  {Jones}(1994)}]{wand1994kernel}%
  \BibitemOpen
  \bibfield  {author} {\bibinfo {author} {\bibfnamefont {M.~P.}\ \bibnamefont
  {Wand}}\ and\ \bibinfo {author} {\bibfnamefont {M.~C.}\ \bibnamefont
  {Jones}},\ }\href@noop {} {\emph {\bibinfo {title} {Kernel smoothing}}}\
  (\bibinfo  {publisher} {CRC press},\ \bibinfo {year} {1994})\BibitemShut
  {NoStop}%
\bibitem [{\citenamefont {Wilson}\ and\ \citenamefont
  {Adams}(2013)}]{wilson2013gaussian}%
  \BibitemOpen
  \bibfield  {author} {\bibinfo {author} {\bibfnamefont {A.}~\bibnamefont
  {Wilson}}\ and\ \bibinfo {author} {\bibfnamefont {R.}~\bibnamefont {Adams}},\
  }\bibfield  {title} {\enquote {\bibinfo {title} {Gaussian process kernels for
  pattern discovery and extrapolation},}\ }in\ \href@noop {} {\emph {\bibinfo
  {booktitle} {International conference on machine learning}}}\ (\bibinfo
  {organization} {PMLR},\ \bibinfo {year} {2013})\ pp.\ \bibinfo {pages}
  {1067--1075}\BibitemShut {NoStop}%
\bibitem [{\citenamefont {Zhang}, \citenamefont {Smith},\ and\ \citenamefont
  {Stechmann}(2021{\natexlab{b}})}]{zhang2021fast}%
  \BibitemOpen
  \bibfield  {author} {\bibinfo {author} {\bibfnamefont {Y.}~\bibnamefont
  {Zhang}}, \bibinfo {author} {\bibfnamefont {L.~M.}\ \bibnamefont {Smith}}, \
  and\ \bibinfo {author} {\bibfnamefont {S.~N.}\ \bibnamefont {Stechmann}},\
  }\bibfield  {title} {\enquote {\bibinfo {title} {Fast-wave averaging with
  phase changes: asymptotics and application to moist atmospheric dynamics},}\
  }\href@noop {} {\bibfield  {journal} {\bibinfo  {journal} {Journal of
  Nonlinear Science}\ }\textbf {\bibinfo {volume} {31}},\ \bibinfo {pages}
  {1--46} (\bibinfo {year} {2021}{\natexlab{b}})}\BibitemShut {NoStop}%
\bibitem [{\citenamefont {Zhang}, \citenamefont {Smith},\ and\ \citenamefont
  {Stechmann}(2022)}]{zhang2022convergence}%
  \BibitemOpen
  \bibfield  {author} {\bibinfo {author} {\bibfnamefont {Y.}~\bibnamefont
  {Zhang}}, \bibinfo {author} {\bibfnamefont {L.~M.}\ \bibnamefont {Smith}}, \
  and\ \bibinfo {author} {\bibfnamefont {S.~N.}\ \bibnamefont {Stechmann}},\
  }\bibfield  {title} {\enquote {\bibinfo {title} {Convergence to precipitating
  quasi-geostrophic equations with phase changes: asymptotics and numerical
  assessment},}\ }\href@noop {} {\bibfield  {journal} {\bibinfo  {journal}
  {Philosophical Transactions of the Royal Society A}\ }\textbf {\bibinfo
  {volume} {380}},\ \bibinfo {pages} {20210030} (\bibinfo {year}
  {2022})}\BibitemShut {NoStop}%
\bibitem [{\citenamefont {Hernandez-Duenas}, \citenamefont {Smith},\ and\
  \citenamefont {Stechmann}(2014)}]{hernandez2014investigation}%
  \BibitemOpen
  \bibfield  {author} {\bibinfo {author} {\bibfnamefont {G.}~\bibnamefont
  {Hernandez-Duenas}}, \bibinfo {author} {\bibfnamefont {L.~M.}\ \bibnamefont
  {Smith}}, \ and\ \bibinfo {author} {\bibfnamefont {S.~N.}\ \bibnamefont
  {Stechmann}},\ }\bibfield  {title} {\enquote {\bibinfo {title} {Investigation
  of boussinesq dynamics using intermediate models based on wave--vortical
  interactions},}\ }\href@noop {} {\bibfield  {journal} {\bibinfo  {journal}
  {Journal of fluid mechanics}\ }\textbf {\bibinfo {volume} {747}},\ \bibinfo
  {pages} {247--287} (\bibinfo {year} {2014})}\BibitemShut {NoStop}%
\bibitem [{\citenamefont {Deng}, \citenamefont {Smith},\ and\ \citenamefont
  {Majda}(2012)}]{deng2012tropical}%
  \BibitemOpen
  \bibfield  {author} {\bibinfo {author} {\bibfnamefont {Q.}~\bibnamefont
  {Deng}}, \bibinfo {author} {\bibfnamefont {L.}~\bibnamefont {Smith}}, \ and\
  \bibinfo {author} {\bibfnamefont {A.}~\bibnamefont {Majda}},\ }\bibfield
  {title} {\enquote {\bibinfo {title} {Tropical cyclogenesis and vertical shear
  in a moist boussinesq model},}\ }\href@noop {} {\bibfield  {journal}
  {\bibinfo  {journal} {Journal of fluid mechanics}\ }\textbf {\bibinfo
  {volume} {706}},\ \bibinfo {pages} {384--412} (\bibinfo {year}
  {2012})}\BibitemShut {NoStop}%
\bibitem [{\citenamefont {Yau}\ and\ \citenamefont
  {Rogers}(1996)}]{yau1996short}%
  \BibitemOpen
  \bibfield  {author} {\bibinfo {author} {\bibfnamefont {M.~K.}\ \bibnamefont
  {Yau}}\ and\ \bibinfo {author} {\bibfnamefont {R.~R.}\ \bibnamefont
  {Rogers}},\ }\href@noop {} {\emph {\bibinfo {title} {A short course in cloud
  physics}}}\ (\bibinfo  {publisher} {Elsevier},\ \bibinfo {year}
  {1996})\BibitemShut {NoStop}%
\bibitem [{\citenamefont {Liu}\ \emph {et~al.}(2023)\citenamefont {Liu},
  \citenamefont {Yau}, \citenamefont {Shima}, \citenamefont {Lu},\ and\
  \citenamefont {Chen}}]{liu2023parameterization}%
  \BibitemOpen
  \bibfield  {author} {\bibinfo {author} {\bibfnamefont {Y.}~\bibnamefont
  {Liu}}, \bibinfo {author} {\bibfnamefont {M.-K.}\ \bibnamefont {Yau}},
  \bibinfo {author} {\bibfnamefont {S.-i.}\ \bibnamefont {Shima}}, \bibinfo
  {author} {\bibfnamefont {C.}~\bibnamefont {Lu}}, \ and\ \bibinfo {author}
  {\bibfnamefont {S.}~\bibnamefont {Chen}},\ }\bibfield  {title} {\enquote
  {\bibinfo {title} {Parameterization and explicit modeling of cloud
  microphysics: Approaches, challenges, and future directions},}\ }\href@noop
  {} {\bibfield  {journal} {\bibinfo  {journal} {Advances in Atmospheric
  Sciences}\ }\textbf {\bibinfo {volume} {40}},\ \bibinfo {pages} {747--790}
  (\bibinfo {year} {2023})}\BibitemShut {NoStop}%
\bibitem [{\citenamefont {Chen}, \citenamefont {Majda},\ and\ \citenamefont
  {Giannakis}(2014)}]{chen2014predicting}%
  \BibitemOpen
  \bibfield  {author} {\bibinfo {author} {\bibfnamefont {N.}~\bibnamefont
  {Chen}}, \bibinfo {author} {\bibfnamefont {A.~J.}\ \bibnamefont {Majda}}, \
  and\ \bibinfo {author} {\bibfnamefont {D.}~\bibnamefont {Giannakis}},\
  }\bibfield  {title} {\enquote {\bibinfo {title} {Predicting the cloud
  patterns of the madden-julian oscillation through a low-order nonlinear
  stochastic model},}\ }\href@noop {} {\bibfield  {journal} {\bibinfo
  {journal} {Geophysical Research Letters}\ }\textbf {\bibinfo {volume} {41}},\
  \bibinfo {pages} {5612--5619} (\bibinfo {year} {2014})}\BibitemShut {NoStop}%
\bibitem [{\citenamefont {Chen}(2023)}]{chen2023stochastic}%
  \BibitemOpen
  \bibfield  {author} {\bibinfo {author} {\bibfnamefont {N.}~\bibnamefont
  {Chen}},\ }\href@noop {} {\emph {\bibinfo {title} {Stochastic methods for
  modeling and predicting complex dynamical systems: uncertainty
  quantification, state estimation, and reduced-order models}}}\ (\bibinfo
  {publisher} {Springer Nature},\ \bibinfo {year} {2023})\BibitemShut {NoStop}%
\bibitem [{\citenamefont {Lopez}(2002)}]{lopez2002implementation}%
  \BibitemOpen
  \bibfield  {author} {\bibinfo {author} {\bibfnamefont {P.}~\bibnamefont
  {Lopez}},\ }\bibfield  {title} {\enquote {\bibinfo {title} {Implementation
  and validation of a new prognostic large-scale cloud and precipitation scheme
  for climate and data-assimilation purposes},}\ }\href@noop {} {\bibfield
  {journal} {\bibinfo  {journal} {Quarterly Journal of the Royal Meteorological
  Society: A journal of the atmospheric sciences, applied meteorology and
  physical oceanography}\ }\textbf {\bibinfo {volume} {128}},\ \bibinfo {pages}
  {229--257} (\bibinfo {year} {2002})}\BibitemShut {NoStop}%
\bibitem [{\citenamefont {Gottschalck}\ \emph {et~al.}(2005)\citenamefont
  {Gottschalck}, \citenamefont {Meng}, \citenamefont {Rodell},\ and\
  \citenamefont {Houser}}]{gottschalck2005analysis}%
  \BibitemOpen
  \bibfield  {author} {\bibinfo {author} {\bibfnamefont {J.}~\bibnamefont
  {Gottschalck}}, \bibinfo {author} {\bibfnamefont {J.}~\bibnamefont {Meng}},
  \bibinfo {author} {\bibfnamefont {M.}~\bibnamefont {Rodell}}, \ and\ \bibinfo
  {author} {\bibfnamefont {P.}~\bibnamefont {Houser}},\ }\bibfield  {title}
  {\enquote {\bibinfo {title} {Analysis of multiple precipitation products and
  preliminary assessment of their impact on global land data assimilation
  system land surface states},}\ }\href@noop {} {\bibfield  {journal} {\bibinfo
   {journal} {Journal of Hydrometeorology}\ }\textbf {\bibinfo {volume} {6}},\
  \bibinfo {pages} {573--598} (\bibinfo {year} {2005})}\BibitemShut {NoStop}%
\end{thebibliography}%

\end{document}